\documentclass{article}

\usepackage{arxiv}

\usepackage[utf8]{inputenc} 
\usepackage[T1]{fontenc}    
\usepackage{hyperref}       
\usepackage{url}            
\usepackage{booktabs}       
\usepackage{amsfonts}       
\usepackage{nicefrac}       
\usepackage{microtype}      
\usepackage{lipsum}

\usepackage{amsmath,amssymb,amsfonts}
\usepackage{amsmath,amsfonts}
\usepackage{algorithm,algpseudocode}
\usepackage{graphicx}
\usepackage{subcaption}
\usepackage{textcomp}
\usepackage{soul,color}

\algnewcommand{\Inputs}[1]{%
	\State \textbf{Inputs:}
	\Statex \hspace*{\algorithmicindent}\parbox[t]{.8\linewidth}{\raggedright #1}
}
\algnewcommand{\Outputs}[1]{%
	\State \textbf{Outputs:}
	\Statex \hspace*{\algorithmicindent}\parbox[t]{.8\linewidth}{\raggedright #1}
}

\title{Data Augmentation Based Malware Detection using Convolutional Neural Networks}

\author{
  Ferhat Ozgur Catak\\
  Simula Research Laboratory\\
  Oslo, Norway\\
  \texttt{ozgur@simula.no} \\
   \And
 Javed Ahmed \\
  Department of Computer Science \\
  Sukkur IBA University, Sukkur, Pakistan\\
  \And
  Kevser Sahinbas \\
  Department of Management Information System\\
  Istanbul Medipol University, Istanbul, Turkey\\
  \And
  Zahid Hussain Khand\\
  Department of Computer Science\\
  Sukkur IBA University, Sukkur, Pakistan
}

\begin{document}
\maketitle

\begin{abstract}
Recently, cyber-attacks have been extensively seen due to the everlasting increase of malware in the cyber world. These attacks cause irreversible damage not only to end-users but also to corporate computer systems. Ransomware attacks such as WannaCry and Petya specifically targets to make critical infrastructures such as airports and rendered operational processes inoperable. Hence, it has attracted increasing attention in terms of volume, versatility, and intricacy. The most important feature of this type of malware is that they change shape as they propagate from one computer to another. Since standard signature-based detection software fails to identify this type of malware because they have different characteristics on each contaminated computer. This paper aims at providing an image augmentation enhanced deep convolutional neural network (CNN) models for the detection of malware families in a metamorphic malware environment. The main contributions of the paper’s model structure consist of three components, including image generation from malware samples, image augmentation, and the last one is classifying the malware families by using a convolutional neural network model.  In the first component, the collected malware samples are converted binary representation to 3-channel images using windowing technique. The second component of the system create the augmented version of the images, and the last component builds a classification model. In this study, five different deep convolutional neural network model for malware family detection is used. 
\end{abstract}


\section{Introduction}
Today, especially in the beginning of the 2000s, motivated by recent promotions of technology, we have started to use many technological devices such as computers, mobile phones and webcams.  Technology manufacturers have started to produce devices that provide users with attractive features. While the development of these features, the issue of security is ignored by developers. Due to these rapid product developments that the developers make for the purpose of releasing the products to the market quickly, many products contain many weaknesses. In terms of malicious software developers, this offers many opportunities.

With the understanding cybersecurity is a vital problem, cyber security activities in organizations have increased. As a result, designing approaches for cyber security activities should be considered. One of the most important cyber security activities is malicious software analysis. Malware is a tool designed for a specific target, often attempting to camouflage itself in another way, with intentions such as file encryption, ransom, preventing a system from working, gaining unauthorized access to a system, data theft, or sabotage. Since, 1046.10 million new malware \footnote{https://www.av-test.org/en/statistics/malware/}  have been found in the first four months of 2020, it is understood that a great deal of effort has been made in the development of malware. Therefore, great efforts are needed to protect against malware attacks.

In order to be effectively protected from malware, the first thing to do is to recognize the malicious software and analyze their behavior well. In this respect, the important point is to identify malicious software and to classify them successfully as well. A family of malicious software also represents the malicious behavior to which it belongs. As a result, the countermeasures to be taken against these behaviors may vary according to the families of malicious software.

Several consecutive operations are generally performed within malware analysis. This task is mainly done using static and dynamic analysis methods including the \texttt{strings} command to get the malicious IP addresses, \texttt{entropy} value if the suspicious executable file, executing the file in an isolated environment to record its behaviour. On the other hand, Malware developers have a broad knowledge of analysis and, with this knowledge, develop a variety of anti-analysis techniques. Anti-debugging and anti-disassembly techniques are the two methods most commonly used by malware developers. Such analysis bypass methods are generally used to produce erroneous results by the disassembler and debugger tools. In anti-debugging methods, malware developers often manipulate pointer address parameters used by jump op-code such as \texttt{jz}, \texttt{jnz}, \texttt{jze}. Anti-debugging techniques are used by developers to ensure that malware samples does not run under a debugger, and in this case, to change the execution flow accordingly. In most cases, the Anti-debugging process will slow down the reverse engineering process.

For this reason, the automated malware detection systems used today do not yield very successful results. Depending on the classification methods used, a malware can be detected as a Trojan using an anti-virus application and another anti-virus application may be labeled as Worm. This has become even more complicated with the advent of sophisticated malware. 

For this reason, especially with the development of machine learning field, this subject is being used in the field of malware analysis. The first use of machine learning algorithms for malicious software analysis is to use API calls as feature vector \cite{8769564}. One of the favorite method performed within the quantification of API calls is n-grams. The main reasons for using n-grams are to give an effort to reduce computation-complexity of the model, to create a simple term-frequency $\times$ inverse-document-frequency (TF-IDF) matrix, and to use traditional algorithms like random forests, decision tree and support vector machine (SVM).

Although such an approach has produced results with high classification performance, they remain inadequate for the current malware infection methods. Malware analysts need sandbox applications to create API call data sets. Basically, a sandbox allows the operation of an isolated virtual machine in a secure, closed network environment, and the malicious software is run and recorded in this running virtual computer. However, malware developers integrate various virtual machine detection code snippets into their malicious code blocks. If they get the impression that they are working on a virtual server or in a sandbox environment, they change behavior to complicate the analysis. These methods used by malware developers are called anti-vm and anti-sandbox, and the most commonly used methods are `Checking CPUID Instruction', `VMWare Magic Number', `Checking for Known Mac Addresses', `Checking for Processes Indicating a VM' , `Checking for Existence of Files Indicating a VM', and `Checking for Running Services'.

Even if they change behavior and block dynamic analysis, the pieces of code are still present, and therefore other machine learning methods can be used to obtain the malware family that they belong to. For this reason, a new approach is used today to create an image of the file and analyze it. The first studies on this subject are generally the creation of a grayscale image of a malware and then the use of classification algorithms.

The main contribution of this research work is to develop a data augmentation enhanced malware family classification model that exploits augmentation for variants of malware clones and takes advantage of convolutional neural network to improve image classification. Herein, we demonstrate that the data augmentation based 3-channel image classification can significantly influence the performance of malware family classification. 

In our previous works \cite{10.7717/peerj-cs.285,DBLP:journals/corr/abs-1905-01999,8806571} we applied a single and two layers LSTM model to detect the malware classes.

The paper is designed as follows: Section \ref{sec:rel} describes the related work. In Section \ref{sec:system}, we provide the system model and consists of two subsections. First subsection presents the image conversion. Second subsection presents the data augmentation. Section \ref{sec:proposed} describes the proposed approach that provides the details of the data augmentation and data enhancement-based CNN malware classification algorithm. Section \ref{sec:exp} provides an extensive analysis of results. Lastly, In Section \ref{sec:conclusion}, we conclude the paper.

\section{Related work}\label{sec:rel}
Malware analysis field has gained considerable attention from research community with rapid development of various techniques for malware detection. There is huge research literature in this area. Since the proposed work is related to image-based analysis using deep learning techniques, the relevant research literature regarding image processing techniques for malware detection are briefly discussed in this section. One of the early studies conducted on malware images was done by Nataraj et. al \cite{10.1145/2016904.2016908}. The authors proposed an image texture analysis-based technique for visualization and classification of different families of malware. This approach converts malware binaries into grayscale images. Malware are classified using K-nearest neighbor technique with Euclidean method. However, the system requires pre-processing of filtering to extract the image texture as features for classification. 

On the other hand, to extract the image texture as features for classification, the system requires pre-processing of filtering. Kesav et al. \cite{6597204} proposed a low-level texture feature extraction technique for malware analysis parallel to Nataraj’s technique. The authors converted malware binaries into images and then extracted discrete wavelets transform based texture features for classification. Aziz et al. \cite{8073489} identify new malware and their variants to extract wavelet transforms-based texture features, and then supply to feed forward artificial neural network for applying classification. Konstantinos et al. \cite{10.1145/3139367.3139400} described a two-step malware variant detection and classification method. In the first step, binary texture analysis applied through GIST. In the second step, these texture features classified by using machine-learning techniques such as classification and clustering to identify malware. Although the works mentioned above \cite{10.1145/2016904.2016908,6597204,8073489,10.1145/3139367.3139400} are helpful to detect and classify new malware and their variants, they still have some limitations. For instance, on the one hand, global texture features lose local information needed for classification. On the another hand, they have significant computation overheads to process a vast amount of malware.

According to Zhang et al. \cite{7823870}, the malware classification problem can be converted into an image classification problem. Their study provides to disassembles executable files into opcode sequences and then convert opcode into images for identifying whether the source file is benign or malware by using CNN. Yue \cite{yue2017imbalanced} presents multifamily malware classification approach by applying convolutional neural network. However, the performance is degraded due to the imbalance of malware families. The author proposes softmax loss function to mitigate this issue. This approach is reactive in nature to deal with scenarios where class imbalance is assumed. 

The other work by Ni et al. \cite{NI2018871} propose a method for malware classification by applying deep learning techniques. Their algorithm uses SimHash and CNN techniques for malware classification. The algorithm converts the malware codes that is disassembled into grayscale images used SimHash algorithm and after that uses convolutional neural network (CNN) to identify their family. The performance improvement is ensured by using some methods such as bilinear interpolation, multi-hash, and major block selection during the process. Cui et al. \cite{8330042} propose a method that applies CNN with the Bat algorithm together in order to robust the accuracy of the model. Their implemented method converts the malicious code into grayscale images. The method’s images are classified by using a convolutional neural network (CNN) and Bat algorithm is used to address the issue of data imbalance among different malware families. The main limitation of this approach is that they used one evaluation criterion to test the model. 

Two stage deep learning neural network is used by Tobiyama et al. \cite{7552276} for infection detection. Initially, the authors generated an image via the extracted behavioral features from the trained recurrent neural network. Later, to classify the feature images, they used CNN. An approach to derive more significant byte sequence in a malware was proposed by Yakura et al. \cite{10.1145/3176258.3176335}. The authors used CNN with attention mechanism to achieve this for the images converted from binaries. MalNet method for malware detection was proposed by Yan et al. \cite{yan2018detecting}. The method automatically learns essential features from the raw data. The method generates grayscale images from opcode sequences. Later, CNN and LSTM are used to learn important features from the grayscale images. Fu et al. \cite{8290767} proposed an approach to visualize malware as an RGB-colored image. Malware classification is performed by merging global and local features using random forest, K-nearest neighbor, and support vector machine. The approach realizes fine-grained malware classification with low computational cost by utilizing the combination of global and local features. Liu et al. \cite{8610223} proposed a malware classification framework based on a bag-of-visual-words (BoVW) model to obtain robust feature descriptors of malware images. The model demonstrates better classification accuracy even for more challenging datasets. The major limitation of this approach is higher computational cost. 

Chen et al. \cite{8703786} conducted an extensive study on the vulnerabilities of the CNN-based malware detectors. The authors proposed two methods to attack recently developed malware detectors. One of these methods achieve attack success rate over 99\% which strongly demonstrates the vulnerability of CNN-based malware detectors. The authors also conducted experiments with pre-detection mechanism to reject adversarial examples and shown its effectiveness in improving the safety and efficiency of malware detectors. Venkatraman et al. \cite{VENKATRAMAN2019377} used similarity mining and deep learning architecture to identify and classify obfuscated malware accurately. The authors used eight different similarity measures to generate similarity matrices and to identify malware family by adopting images of distance scores. The advantage of this approach is that it requires less computational cost as compared to classical machine learning based methods. Dai et al. \cite{DAI201830} proposed a malware detection method using hardware features due to inherent deficiencies in software methods. The approach dumps the malware memory of runtime to binary files, then grayscale image is extracted from the binary files. A fixed size images are generated from the grayscale image and histogram of gradient is used to extract image features. Finally, malware classification is done using the popular classifier algorithms. One of the limitations for this approach is that it cannot provide against fileless malware. Gibert et al. \cite{gibert2019using} propose a file agnostic deep learning approach for malware classification. The malicious software are grouped into families based on a set of discriminant patterns extracted from their visualization as images. Yoo et al. \cite{8963615} propose multiclass CNN model to classify exploit kits. On of the root of malware contamination are exploit kits. This type of attack has rapidly increased and detection rate is quite low. The authors proposed limited grayscale, size-based hybrid model, and recursive image update method to enhance classification accuracy.

Traditional machine learning methods are applied in most of the existing state of the art. Our study uses a deep learning method and differs from most other studies examined in this section. Deep learning methods are not algorithmically new and easy to implement. They can be trained with high-performance computations on systems such as GPUs. Today, they have become prevalent in the field of machine learning. Some of the studies examined also used deep learning methods, but our approach differs from these studies because we used five different deep convolutional neural network models for malware family classification. It is evident from the results that 3-channel image classification can significantly influence malware family detection's performance. The main contribution that makes this study stand out regarding the existing state of art examined in this section is applying data augmentation enhanced malware family classification model. This model exploits augmentation for variants of malware clones and take advantage of CNN to improve image classification.

\section{System model}\label{sec:system}
The system architecture of the proposed model is composed into three different components. The first component is image conversion of malware samples using decimal representation and entropy values of each byte. The second component is image augmentation component. The last one is CNN based malware family classification.

\subsection{Image conversion}

\begin{figure}[htbp!]
	\centering
	\includegraphics[width=1.0\linewidth]{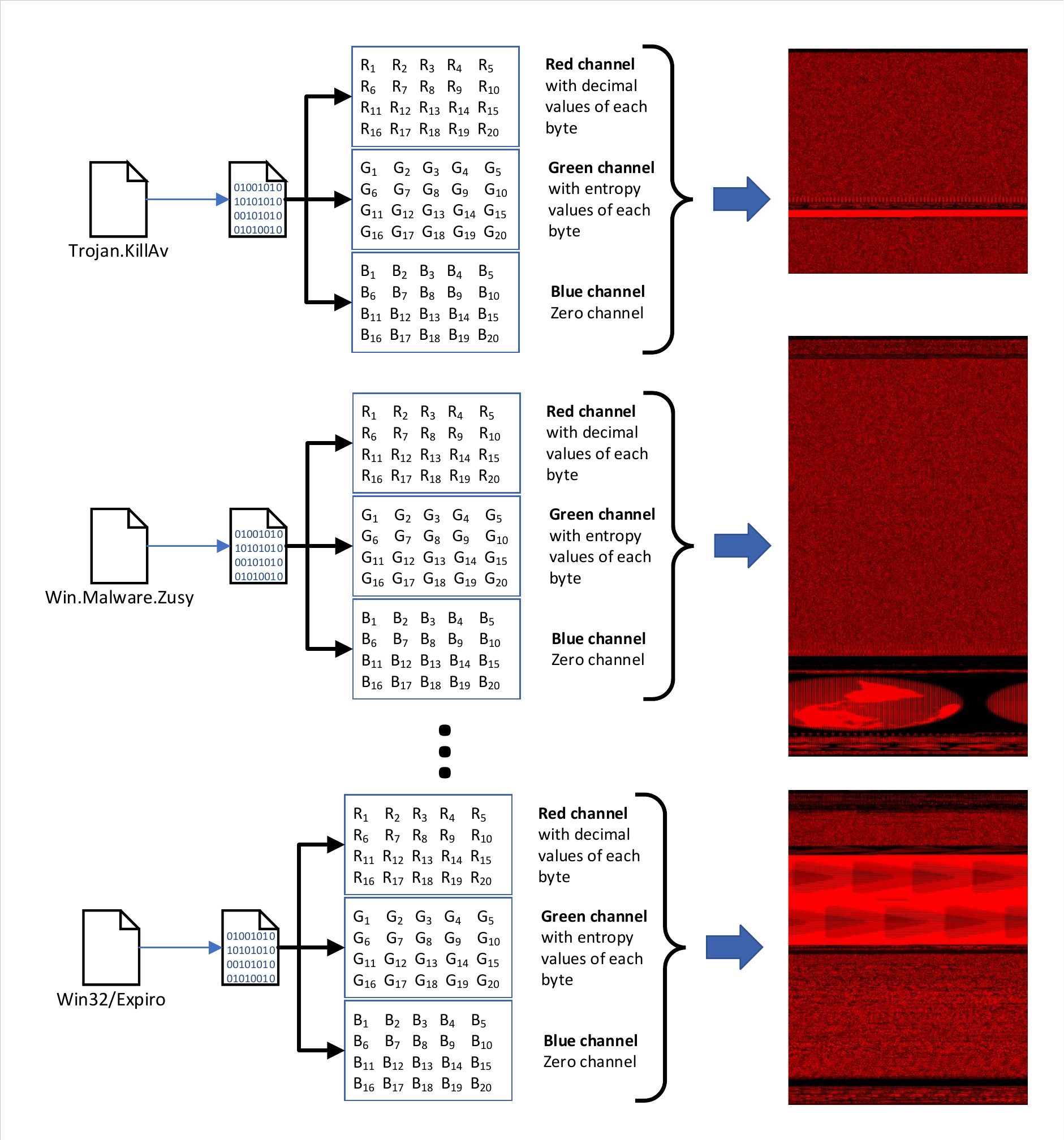}
	\caption{The architecture of the proposed 3-channel image representation of malware samples. Given input malware samples, RGB representations are computed by applying as explained in Section \ref{sec:basic}.}
	\label{fig:img-conv}
\end{figure}

Malware samples collected from Github platform, in the first step, are labeled using \textit{ClamAV} open source command-line antivirus software. The model architecture is illustrated in Figure \ref{fig:img-conv}.


The input of the first component of the malware detection system is a collection of malware stored in different formats such as portable executable, Word, PDF. These malware are then converted into 3-channels PNG files as shown in Figure \ref{fig:img-conv}.

%

\subsection{Data Augmentation}

The key problem with malware detection model is data diversity. There are many alternative methods are available for solving these problems. One approach to solve this problem involves the use of data augmentation. Data augmentation can be defined as a strategy to artificially increase the variety of input instances for training phase, without really collecting new instances. 

Additive noise is the most used technique for data augmentation to build reliable machine learning models. Gaussian, Laplacian and Poisson noises are the most used techniques to enhance the input dataset. Laplacian noise is derived eventually from white (Gaussian) noise \cite{hida2008lectures}.

\subsubsection{Additive Gaussian}
Additive Gaussian noise is a fundamental noise model used in information theory to simulate the impact of many random methods that happen in nature. The Additive Gaussian noise flow is represented by a series of outputs $Y_i$ at a discrete-time event index $i$. $Y_i$ is the sum of the input $X_i$ and noise, $Z_i$, where $Z_i$ is independent and identically distributed and picked from a zero-mean normal distribution, including variance $N$. The $Z_i$ are further assumed to not be correlated with the $X_i$.

\begin{equation}
\begin{aligned}
Z_i & \sim \mathcal{N}(0, N) \\
Y_i & = X_i + Z_i
\end{aligned}
\end{equation}

\subsubsection{Additive Poisson}
Poisson noise is a kind of noise that can be represented by a Poisson process. A discrete random variable $X$ is said to have a Poisson distribution with parameter $\lambda > 0$ , if, for $k = 0, 1, 2, \cdots,$ the probability mass function of $X$ is given by:
\begin{equation}
f(k;\lambda )=\Pr(X=k)={\frac {\lambda ^{k}e^{-\lambda }}{k!}}
\end{equation}

where $e$ is Euler's number, and $k!$ is the factorial of $k$.
\subsubsection{Additive Laplace}
The Laplace distribution is a continuous probability distribution that sometimes described the double exponential distribution because it can be considered as two exponential distributions with an extra location parameter joined together. 

A random variable has a $\textrm{Laplace}$ distribution if its probability density function is

\begin{equation}
\begin{aligned}
f(x\mid\mu,b) & = \frac{1}{2b} \exp \left( -\frac{|x-\mu|}{b} \right) \\
& = \frac{1}{2b}
\left\{\begin{matrix}
\exp \left( -\frac{\mu-x}{b} \right) & \text{if }x < \mu
\\[8pt]
\exp \left( -\frac{x-\mu}{b} \right) & \text{if }x \geq \mu
\end{matrix}\right.
\end{aligned}
\end{equation}

\section{Preliminaries} \label{sec:mal-dev-tools}
Malware developers try to hide the malicious code snippets they place on legitimate software from malware analysts and antivirus programs using different methods. In addition, malware software developers use codes and frameworks that belong to malware families that perform similar malicious activities, rather than rebuilding malware code fragments. For this reason, when these malware are converted into a executable file (example: PE for Windows) to be suitable for the target platform on which they will be run, they are very similar when binary analysis is performed. The signature-based security components used today are very vulnerable to changes in the code, which reduces their detection capabilities. Developers generally use two different methods to replace the malicious code content when contaminated software infects from one host computer to another computer; \textit{polymorphic} and \textit{metamorphic} malware.

\begin{figure}[h]
	\centering
	\includegraphics[width=1.0\linewidth]{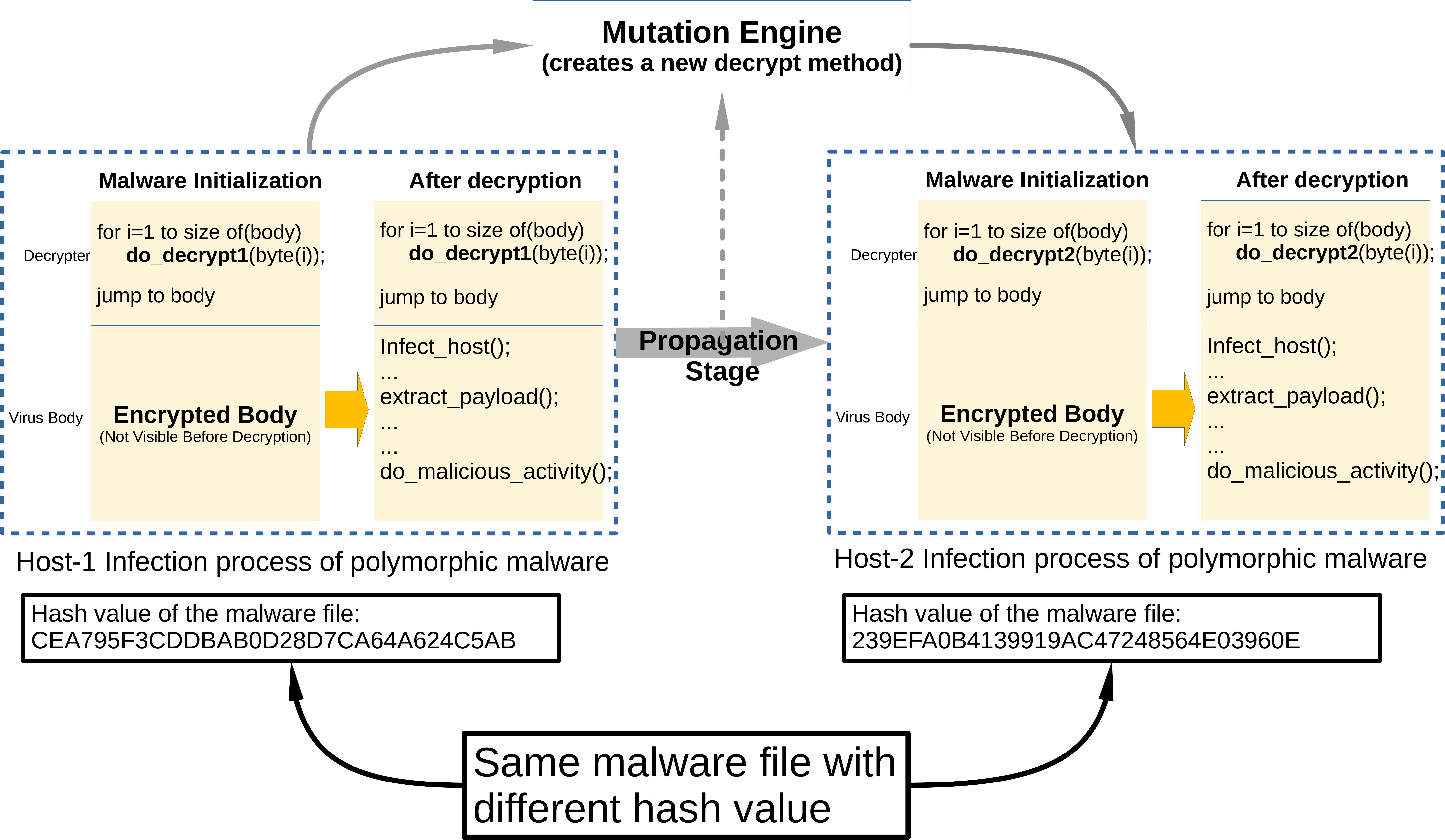}
	\caption{Typical polymorphic malware propagation}
	\label{fig:polymorphic}
\end{figure}

Polymorphic malware changes the code during each infection and changes its appearance, but its malicious activity remains the same. \cite{drew2017polymorphic}. The Polymorphic malware uses different code obfuscation techniques while transmitting from one host to another host. Their approaches are encryption of malicious code parts, obfuscating variable names using character code shifts, register reassignments, equivalent code replacements, and removal of white space or code minification. Figure \ref{fig:polymorphic} shows the components and phases of a typical polymorphic malware during infection. An infected host is shown on the left side of the figure. Infection occurs in 2 stages; initialization and decryption. In the first stage, the malware is working to behave normally on the host. In the second stage, the obfuscated malware payload will open with the decrypter component, and then the \textit{jump} command is used to change the stack on the address to the new malware payload location on memory. Polymorphic malware will change its appearance by using mutation engine with obfuscation techniques while passing from a host to another host. As a result, it will actually perform the same harmful activity, but its appearance will be different, and as a result infected executable files with different hash values will be created.

\begin{figure}[h]
	\centering
	\includegraphics[width=1.0\linewidth]{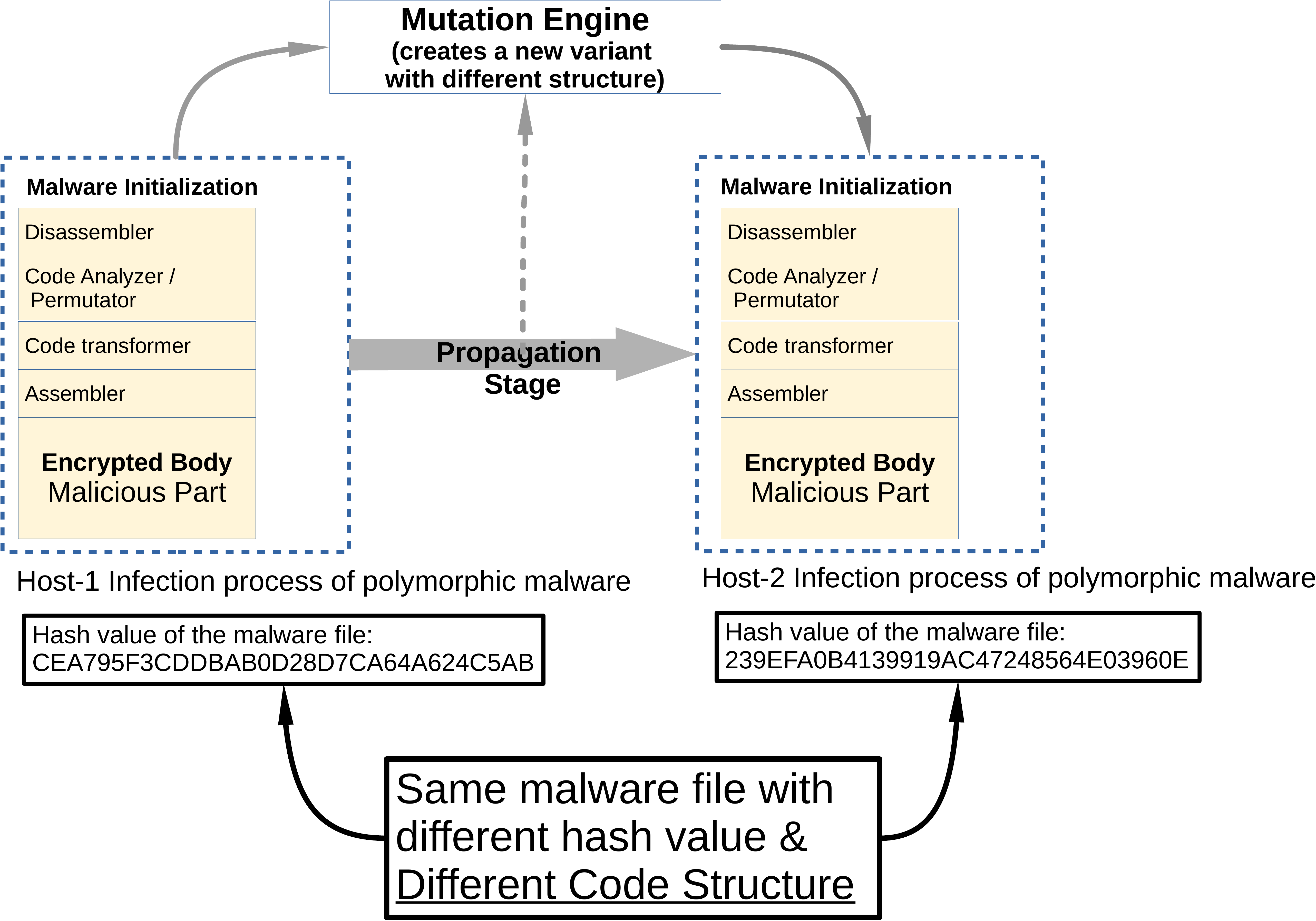}
	\caption{Typical metamorphic malware propagation}
	\label{fig:metamorphic}
\end{figure}

In Metamorphic malware, the situation is a bit more complicated. Although the obfuscation techniques are applied in the same way, this time the code flux is changed. As seen in Figure \ref{fig:metamorphic}, a typical metamorphic malware has more components and its structure has become more complex. This time malware has different components such as disassembler, code analyzer / permutator, code transformer, assembler, and malicious payload. 

\section{Proposed approach}\label{sec:proposed}

This section presents the results of the data augmentation, data enhancement-based CNN malware classification algorithm. The basic idea of Augmented-CNN based malware classification techniques is introduced in Section \ref{sec:basic}. The implementation of the porposed technique is described in Section \ref{sec:impl}.

\subsection{Basic idea}\label{sec:basic}
As previously mentioned in Section \ref{sec:mal-dev-tools}, malware developers are trying to evade security components using different methods. These methods are usually in the form of adding noise to the executable files' binary form. One of the areas dealing with noisy data is the image classification task. One of the methods used to overcome this problem and to classify images from different angles in a more reliable way is the image augmentation technique. As part of this study, malware samples have been converted to 3-channel images. The evasion techniques that malware developers have added are reflected in these images as noise. We used image augmentation techniques in this study so that the noise in the images does not affect the classification performance.

\begin{figure}[h]
	\centering
	\begin{subfigure}{0.20\linewidth}
		\includegraphics[width=\linewidth]{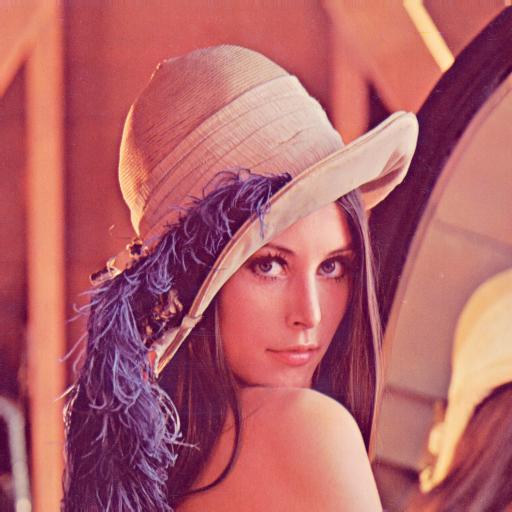} 
		\label{fig:1a}
	\end{subfigure}\hfill
	\begin{subfigure}{0.20\linewidth}
		\includegraphics[width=\linewidth]{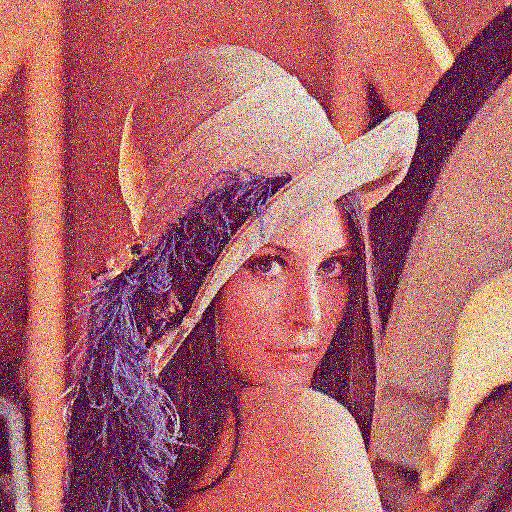}
		\label{fig:1b}
	\end{subfigure}\hfill	
	\begin{subfigure}{0.20\linewidth}
		\includegraphics[width=\linewidth]{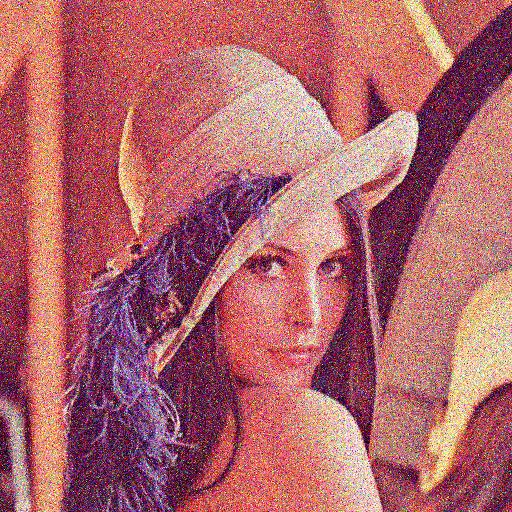}
		\label{fig:1c}
	\end{subfigure}\hfill
	\begin{subfigure}{0.20\linewidth}
		\includegraphics[width=\linewidth]{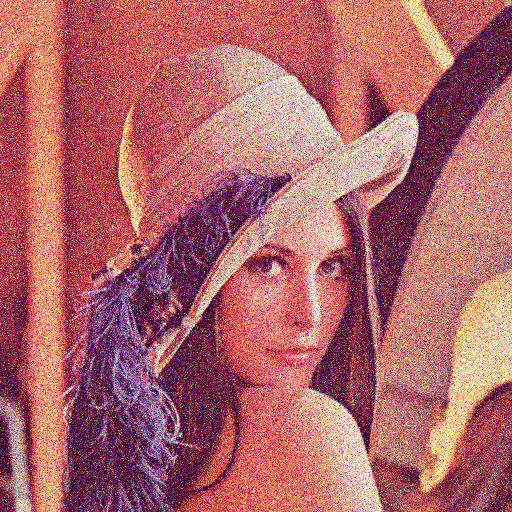}
		\label{fig:1d}
	\end{subfigure}\hfill
	\begin{subfigure}{0.20\linewidth}
		\includegraphics[width=\linewidth]{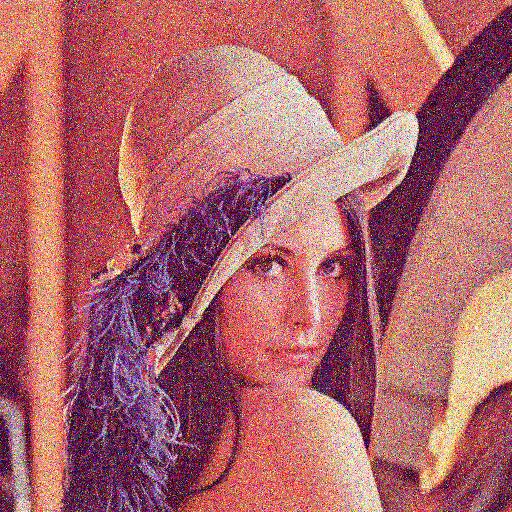}
		\label{fig:1e}
	\end{subfigure}
	\caption{This figure shows the different additive laplace noise to original Lena image.}
	\label{fig:lena-laplace}
\end{figure}

In Figure \ref{fig:lena-laplace}, new images are created when different laplace noises are added to the original image.

\begin{figure}[h]
	\centering
	\begin{subfigure}{0.19\linewidth}
		\includegraphics[width=\linewidth]{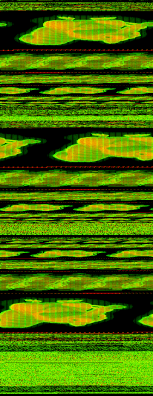} 
		\label{fig:1a}
	\end{subfigure}\hfill
	\begin{subfigure}{0.19\linewidth}
		\includegraphics[width=\linewidth]{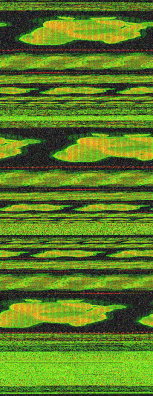}
		\label{fig:1b}
	\end{subfigure}\hfill	
	\begin{subfigure}{0.19\linewidth}
		\includegraphics[width=\linewidth]{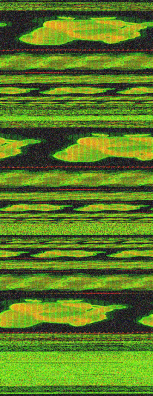}
		\label{fig:1c}
	\end{subfigure}\hfill
	\begin{subfigure}{0.19\linewidth}
		\includegraphics[width=\linewidth]{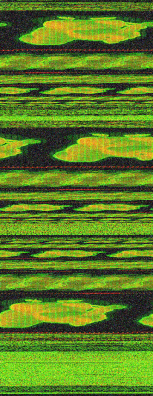}
		\label{fig:1d}
	\end{subfigure}\hfill
	\begin{subfigure}{0.19\linewidth}
		\includegraphics[width=\linewidth]{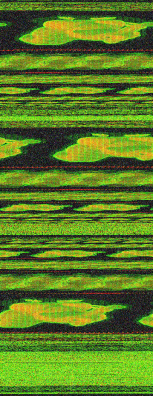}
		\label{fig:1e}
	\end{subfigure}
	\caption{This figure shows the different additive laplace noise to \textit{Trojan/Win32.VBKrypt.C122300} malware.}
	\label{fig:malware-laplace}
\end{figure}

We used the \textit{imgaug} Python library for implementation and increased our dataset to 5 times using \textit{AdditiveGaussian}, \textit{AdditiveLaplace} and \textit{AdditivePoisson} noise addition methods. In Figure \ref{fig:malware-laplace}, new images are created with different laplace noises for \textit{Trojan/Win32.VBKrypt.C122300} malware.

Our main tasks are to enhance data using data augmentation and classify malware samples according their family using malware images based CNN model. Malware images' basic idea is create multi-channel images using byte streams and entropy values of each 8-bits streams. Table \ref{tbl:common-notations} presents notations to evaluate the malware classifier model performance and the commonly used variables is presented for convenience.

\begin{table}[h]
	\caption{Commonly used variables and notations}
	\label{tbl:common-notations}
	\begin{tabular}{ll}
		\hline
		Variables/notations & Description \\ \hline
		$\mathcal{X}$ & Original input dataset \\
		$\mathcal{X}_{aug}$ & Augmednted version of input dataset $X$ \\
		$f_{aug}^{m}$ & Augmentation function $m$ \\
		$\epsilon$ & Augmentation threshold \\
		Acc & Accuracy of the classifier \\
		$k$ & Number of classes \\
		$t$ & Number of augmentation functions \\
		\hline
	\end{tabular}
\end{table}

\subsection{Analysis of the proposed algorithm}
The reason behind of this study is the idea that using the law of large numbers theory, we have opportunity to obtain more accurate classifier model (for this work malware classification) by creating new samples that is comparable to original models which are created with original input instances. 

In the proposed approach, there is a set of augmentation functions that acts a data creation source for CNN model. The single augmentation function, $f_{aug}^{m}$, is defined as follows:
\begin{equation}
\mathcal{X}_{aug}^{(m)} = f_{aug}^{m}(\mathcal{X})
\end{equation}

The each augmented dataset, $\mathcal{X}_{aug}^{(m)}$, using each augmentation algorithm, $f_{aug}^{(m)}$, is combined into a single enhanced dataset. The final augmented dataset is defined as follows:

\begin{equation}
\mathcal{X}_{aug} = \bigcup_{i=1}^t \mathcal{X}_{aug}^{(i)}
\end{equation}

where $t$ is the number of augmented dataset, $\mathcal{X}_{aug}^{(i)}$ is the $i$th augmented dataset.

\subsection{Implementation of the model}\label{sec:impl}

The pseudocode of transformation of PE executable to multichannel images  is shown in Algorithm \ref{alg:imageconv}. The each member ($e^{(i)}$) of collected Windows executable file set, $\mathcal{E}$, is converted multi-channel images in lines 5-6. For the first channel of the executable, one byte is read and then converted to the decimal representation in line 5. The decimal value is assigned to the first channel of the corresponding pixel,  $\mathcal{R}(i,j,0)$. In the same way, this byte's entropy value is assigned to the second channel of the corresponding pixel,  $\mathcal{R}(i,j,1)$. We used \texttt{imgaug} library which uses 3-channel PNG images as input. On the other hand, we created 2-channel PNG images in this research.  Since the \texttt{imgaug} software library requires three channels images, we had to fill the last channel, the Blue channel, with zeros.

\begin{algorithm}[h]
	\caption{PE malware to image conversion}\label{alg:imageconv}
	\begin{algorithmic}[1]
		\Inputs{PE executable set $\mathcal{E}$, image width $w$, image height $h$, channel size $c$}
		\For{each $e^{(i)} \in \mathcal{E}$}
		\State $\mathcal{R} \gets zeros(w,h,c)$ where $\mathcal{R} \in \mathbb{R}^{w \times h \times c }$
		\Comment Create a zero filled matrix
		\For{each byte value $b^{(j)} \in e^{(i)}$}
		\State $\mathcal{R}(i,j,0) \gets decimal(b^{(j)})$
		\Comment 1st channel with value $\in [0,255]$
		\State $\mathcal{R}(i,j,1)  \gets -\sum_{x \in b^{(j)}}\left(p(x) \cdot \log p(x) \right) $
		\Comment 2nd channel with entropy $\in [0,255]$
		\EndFor
		\EndFor
		\Outputs{Image dataset $\mathcal{X}$}
	\end{algorithmic}
\end{algorithm}

The pseudocode of data-augmentation enhanced CNN malware detection are shown in Algorithm \ref{alg:aug}. The augmentation procedure is implemented based on random noise assigment of each channel of the training dataset, $\mathcal{X}$, with a set of augmentation functions, $F_{aug}$.

\begin{algorithm}[h]
	\caption{Data enhancement}\label{alg:aug}
	\begin{algorithmic}[1]
		\Inputs{$\mathcal{X}=\{\{(\mathbf{x}_i, y_i)\mid i=1,...,n\},\mathbf{x}_i \in \mathbb{R}^p,\, y_i \in \{-1, +1\}\}_{i=1}^m$, Augmentation function set $F_{aug}$}
		\State Initialize $\mathcal{X}_{aug}^{(i)}=\mathcal{X}$
		\For{each $f_{aug}^{(i)} \in F_{aug}$}
		\State $\mathcal{X}_{aug}^{(i)} \gets f_{aug}^{(i)}(\mathcal{X})$
		\State $\mathcal{X} \gets \mathcal{X} \bigcup \mathcal{X}_{aug}^{(i)}$
		\EndFor
		\Outputs{Enhanced dataset $\mathcal{X}$}.
	\end{algorithmic}
\end{algorithm}

\section{Experiments}\label{sec:exp}
In this section, we use our public malware dataset \footnote{https://github.com/ocatak/malware\_api\_class} that can be accessed publicly. The malware classification model is compared with the original dataset. In Section \ref{sec:dataset-detail}, we explain the dataset and parameters that are used in our experiments. The conventional CNN is applied the dataset and we find the classification performance in Section \ref{sec:dataset-conv-cnn}. In Section \ref{sec:dataset-proposed}, we show the emprical results of proposed augmented CNN training algorithm.

\subsection{Experimental setup}\label{sec:exp-setup}

To our knowledge, there is no public benchmark dataset for malware images approach to make an evaluation comparison. We apply our dataset with different hyper-parameters to indicate the effectiveness and classification performance of the proposed model.

The experiments are done using the Python programming language and machine learning libraries Keras, Tensorflow, and Scikit-learn. We used the Keras library to build CNN networks. 

\subsection{Dataset detail}\label{sec:dataset-detail}
We trained our classifiers with our public dataset which is summarized in Table \ref{tbl:ds-desc} with seven different classes including Worm, Downloader, Spyware, Adware, Exploit, Malware and Benign. 

\begin{table}[h]
	\centering
	\caption{Description of the training dataset used in the experiments}
	\label{tbl:ds-desc}
	\begin{tabular}{lr}
		\hline
		\textbf{Malware type} & \textbf{\#Inst.} \\
		\hline
		Worm & 1620 \\
		Downloader & 1512 \\
		Spyware & 582 \\
		Adware & 1146 \\
		Exploit & 138 \\
		Malware & 456 \\
		Benign & 308 \\ \hline
		\textbf{Total} & \textbf{5762} \\ \hline
	\end{tabular}
\end{table}

There are 5762 malware samples from different classes in this dataset. The Cuckoo Sandbox application, as explained above, is used to obtain the Windows API call sequences of malicious software, and VirusTotal Service is used to detect the classes of malware.

\begin{figure}[h]
	\centering
	\includegraphics[width=0.7\linewidth]{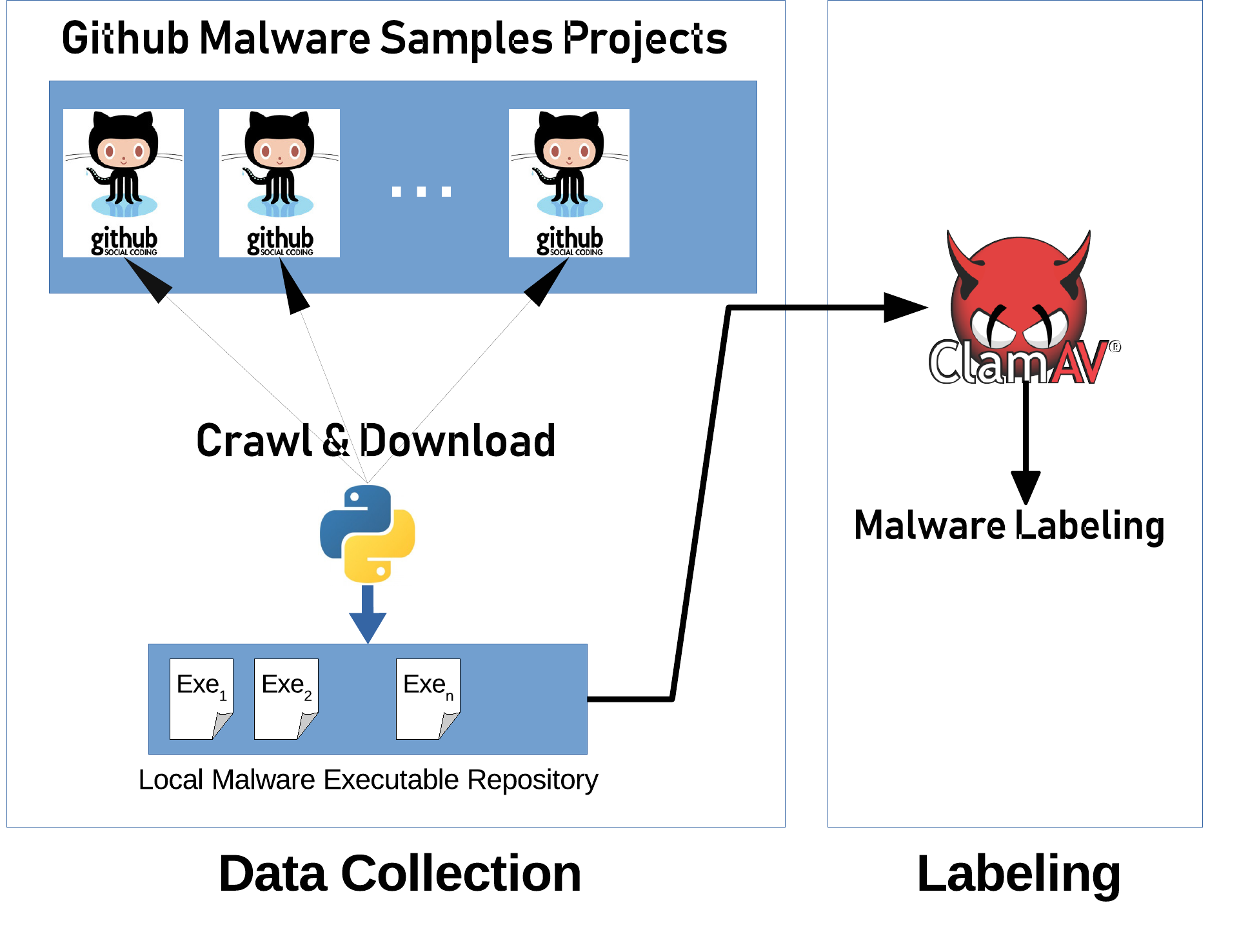}
	\caption{General system architecture. Architecture consists of 3 parts; data collection, Data pre-processing, and Data classification.}
	\label{fig:dataset}
\end{figure}

Figure \ref{fig:dataset} illustrates the system architecture used to collect the data and labeling process. Our system consists of two main parts, data collection, and labeling.

\subsection{Evaluation}\label{sec:eval}
Although the dataset that is applied in our method is almost balanced, performance evaluation in terms of traditional accuracy not sufficient to obtain an optimal classifier. Besides, we apply four metrics such as the overall prediction accuracy, average recall, average precision \cite{Turpin:2006:UPV:1148170.1148176} and F1-score, to estimate the classification accuracy that are used as measurement metrics in machine learning common \cite{Manning:2008:IIR:1394399,Makhoul99performancemeasures}.

Precision is the ratio of predicted positive classes to positive predictions. Precision is estimated in Eq. \ref{eqn:prec}. 
\begin{equation}
\label{eqn:prec}
Precision = \frac{Correct}{Correct + False}
\end{equation}

Recall is the ratio of positive classes to the sum of positive correct estimation and false negative. It can be called Sensitivity. Recall is indicated in Eq. \ref{eqn:recall}.
\begin{equation}
\label{eqn:recall}
Precision = \frac{Correct}{Correct + Missed}
\end{equation}

First, our proposed evaluation model estimates precision and recall for each and then calculate their mean. In Eq. \ref{eqn:avgprec} and Eq. \ref{eqn:avgrecall}, we present average precision and recall. 
\begin{equation}
\label{eqn:avgprec}
Precision_{avg} = \frac{1}{n_{classes}}\sum_{i=0}^{n_{classes}-1}{\left(Prec_i \times num\_of\_instances_i \right) }
\end{equation}
\begin{equation}
\label{eqn:avgrecall}
Recall_{avg} = \frac{1}{n_{classes}}\sum_{i=0}^{n_{classes}-1}{\left(Recall_i \times num\_of\_instances_i \right) }
\end{equation}
The average precision and recall values are calculated using the multiplication of recall and the number of instance in the corresponding class. Precision and Recall are evaluated together in F-measure. It is the harmonic mean of precision and recall. F-measure is provided in Eq. \ref{eqn:fmeasure}.

\begin{equation}
\label{eqn:fmeasure}
F_1 = 2 \times \frac{Prec_{avg} \times Recall_{avg}}{Prec_{avg} + Recall_{avg}}
\end{equation}

\subsection{Dataset results with conventional CNN}\label{sec:dataset-conv-cnn}

\begin{figure}[h]
	\centering
	\includegraphics[width=0.5\linewidth]{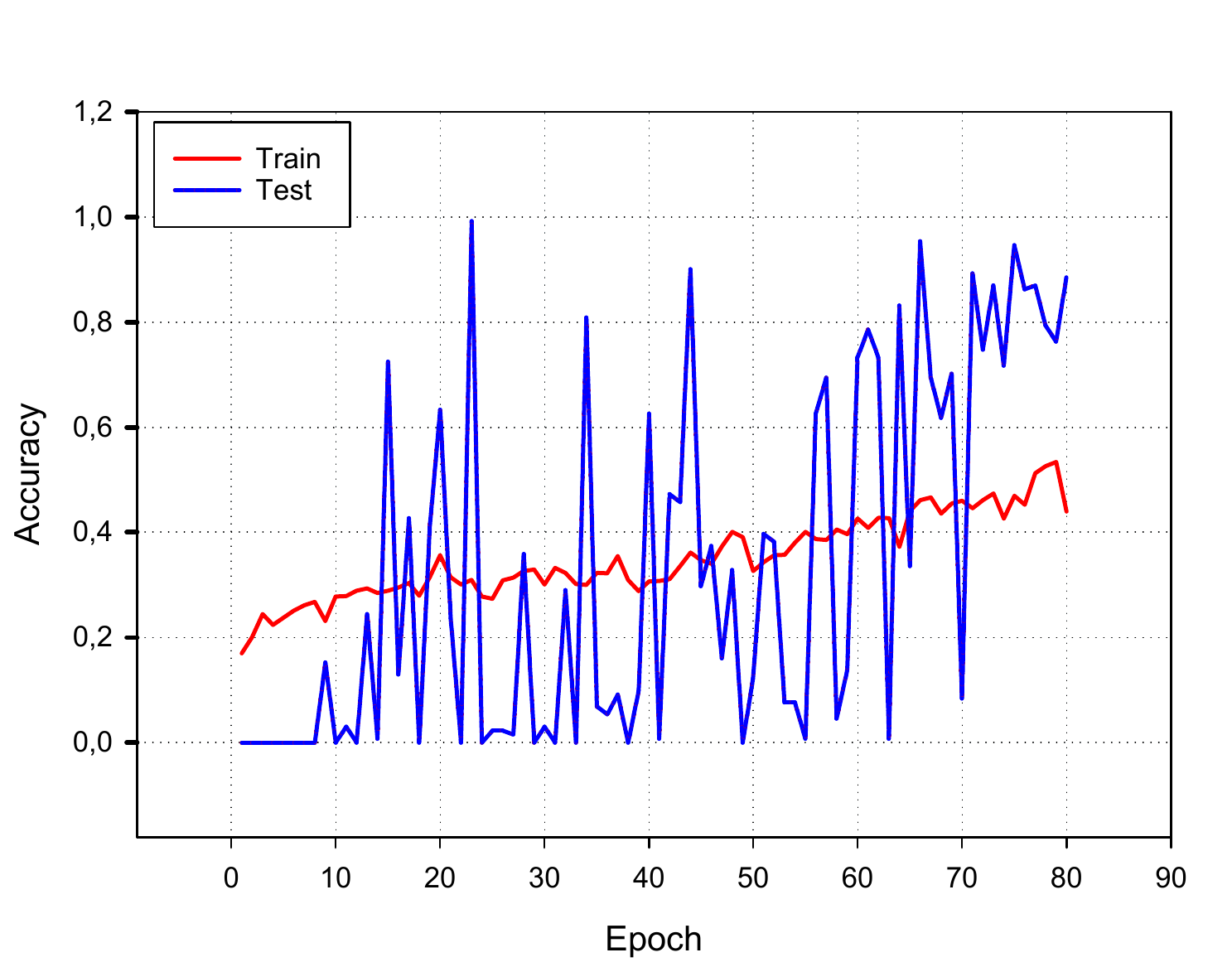}
	\caption{The figure shows the accuracy changes over learning iterations. As can be seen, although the training dataset shows more stable progress, the test dataset is less stable, although it progresses together.}
	\label{fig:org-acc}
\end{figure}

\begin{figure}[h]
	\centering
	\includegraphics[width=0.5\linewidth]{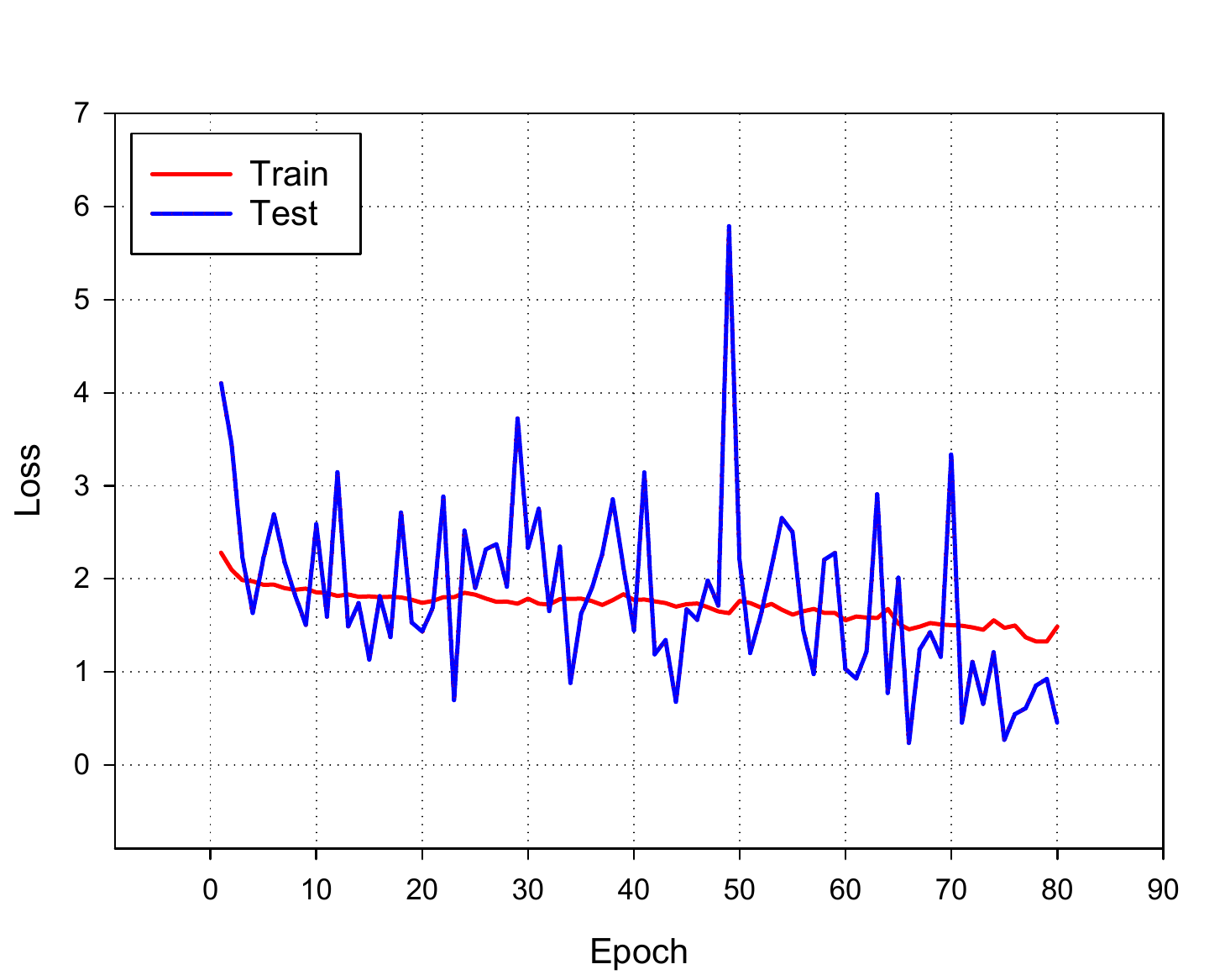}
	\caption{The figure shows the loss changes over learning iterations. As can be seen, although the training dataset shows more stable progress, the test dataset is less stable, although it progresses together like in Figure \ref{fig:org-acc}.}
	\label{fig:org-loss}
\end{figure}

Figure \ref{fig:org-acc} presents the accuracy performance of the conventional CNN model for our experimental data set. As shown in figure, the model becomes its steady state after 80th epoch. Also, Figure \ref{fig:org-loss} shows the loss value changes of classification model through epochs.

\begin{figure}[h]
	\centering
	\includegraphics[width=0.5\linewidth]{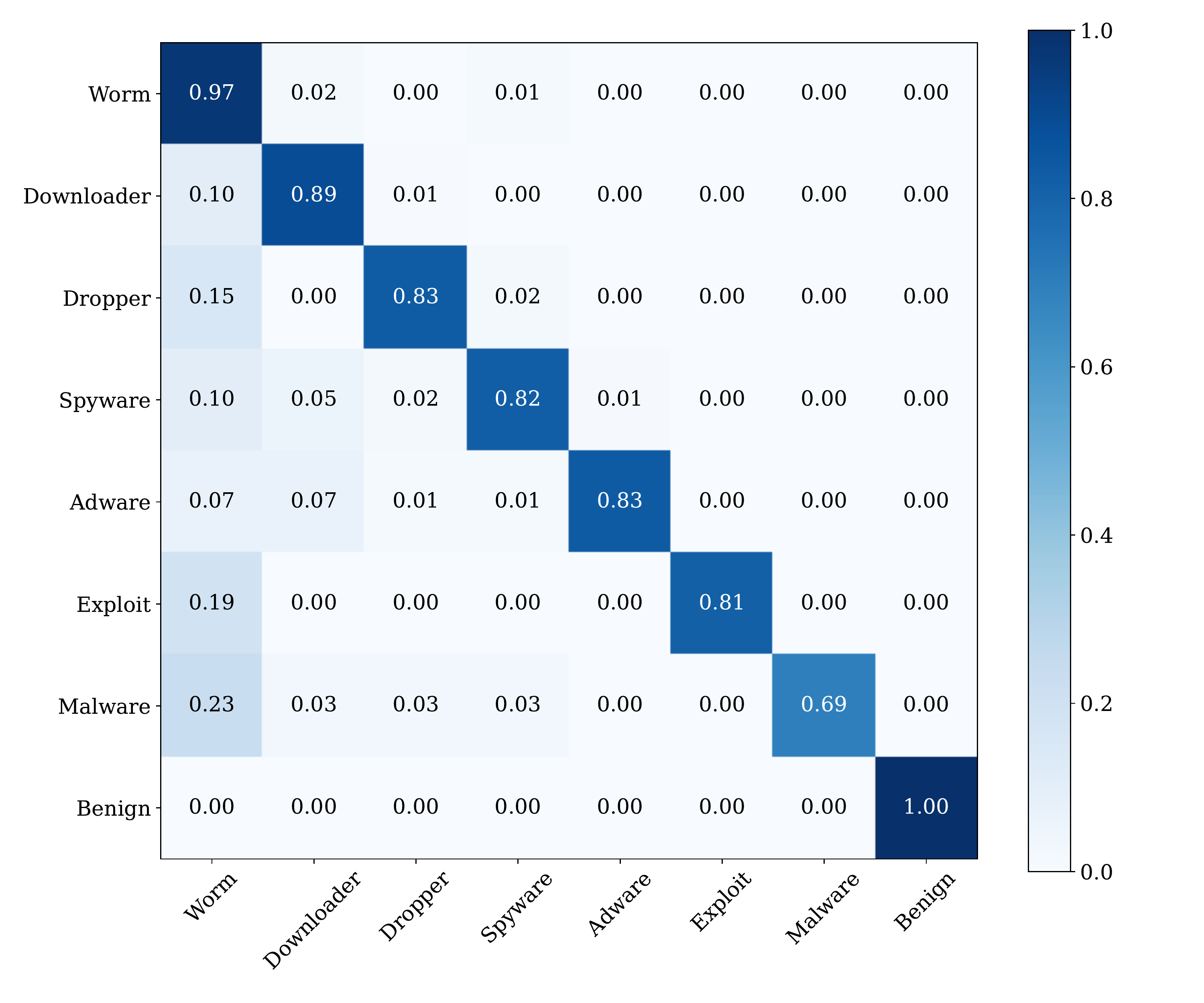}
	\caption{The confusion matrix of the CNN model, which was trained using the original dataset.}
	\label{fig:org-conf-mat}
\end{figure}

A confusion matrix is applied to evaluate the performance of our model. The findings from Figure \ref{fig:org-conf-mat} show the confusion matrix that was trained by using the original dataset by using CNN model. The findings of the confusion matrix indicate that the classification model performance is not good enough for the malware detection.

The testing classification performance is measured through accuracy, precision, recall and $F_1$ measure. Table \ref{tbl:org-class-report} shows the best performance of the conventional CNN method of each malware family.

\begin{table}[htbp!]
	\caption{Classification report of conventional CNN for each malware class}
	\label{tbl:org-class-report}
	\centering
	\begin{tabular}{r|ccc}
		\hline
		& \textbf{Precision} & \textbf{Recall} & $\mathbf{F_1}$ \\ \hline \hline
		Worm & 0.60 & 0.58 & 0.59 \\
		Downloader & 0.82 & 0.11 & 0.20 \\
		Dropper & 0.62 & 0.05 & 0.10 \\
		Spyware & 0.39 & 0.69 & 0.50 \\
		Adware & 0.22 & 0.72 & 0.34 \\
		Exploit & 0.86 & 0.26 & 0.40 \\
		Malware & 0.00 & 0.00 & 0.00 \\
		Benign & 0.77 & 0.83 & 0.80 \\ \hline
	\end{tabular}
\end{table}

As can be seen from the confusion matrix and classification report, the classification performance of the model obtained with conventional CNN is rather low. According to these results, a standard CNN model with RGB type 3-channel image training dataset is not suitable for malware detection and classification.

\subsection{Dataset results with proposed method}\label{sec:dataset-proposed}

\begin{figure}[H]
	\centering
	\begin{subfigure}{0.49\linewidth}
		\includegraphics[width=\linewidth]{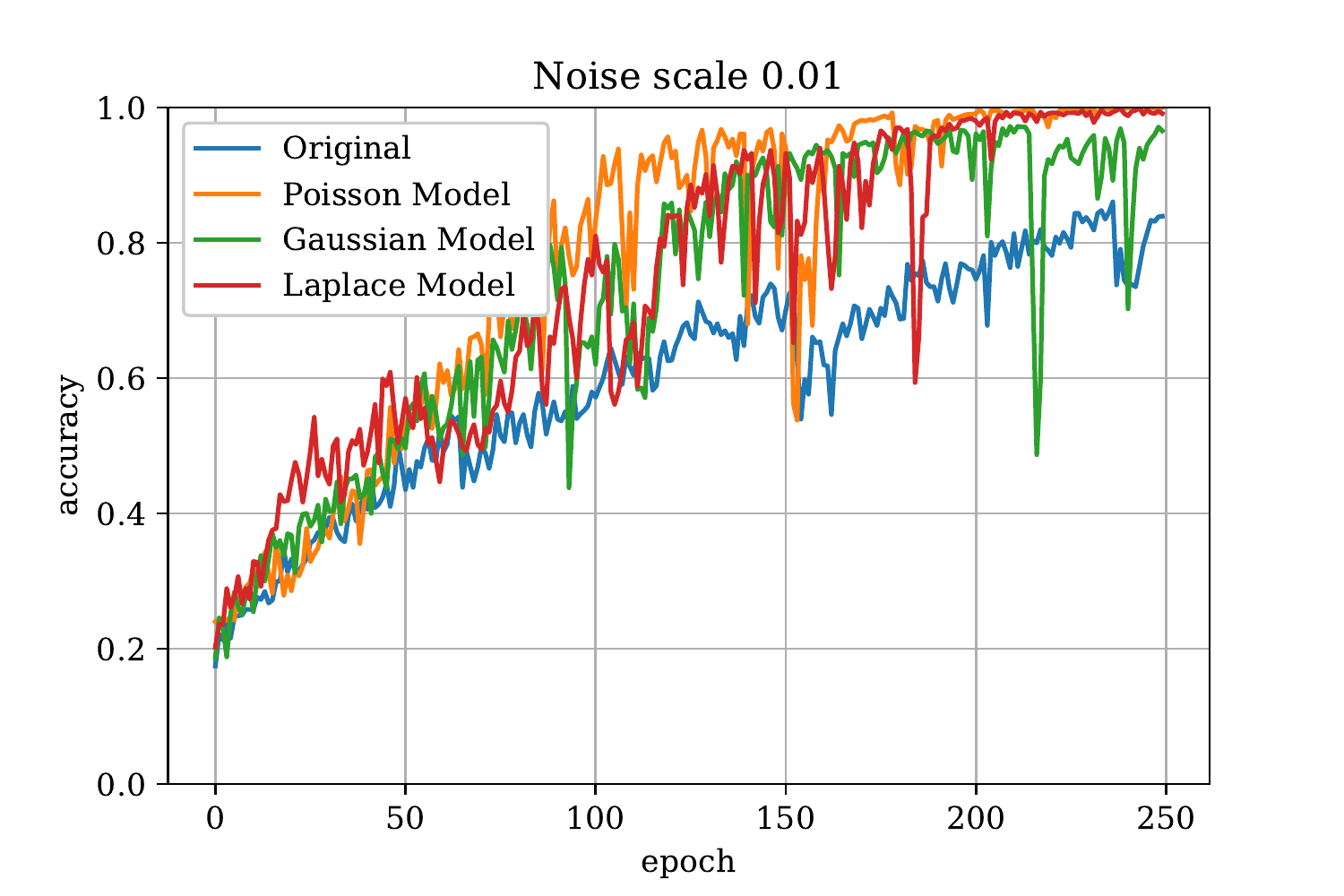} 
		\label{fig:acc-plt-0.01}
	\end{subfigure}\hfill
	\begin{subfigure}{0.49\linewidth}
		\includegraphics[width=\linewidth]{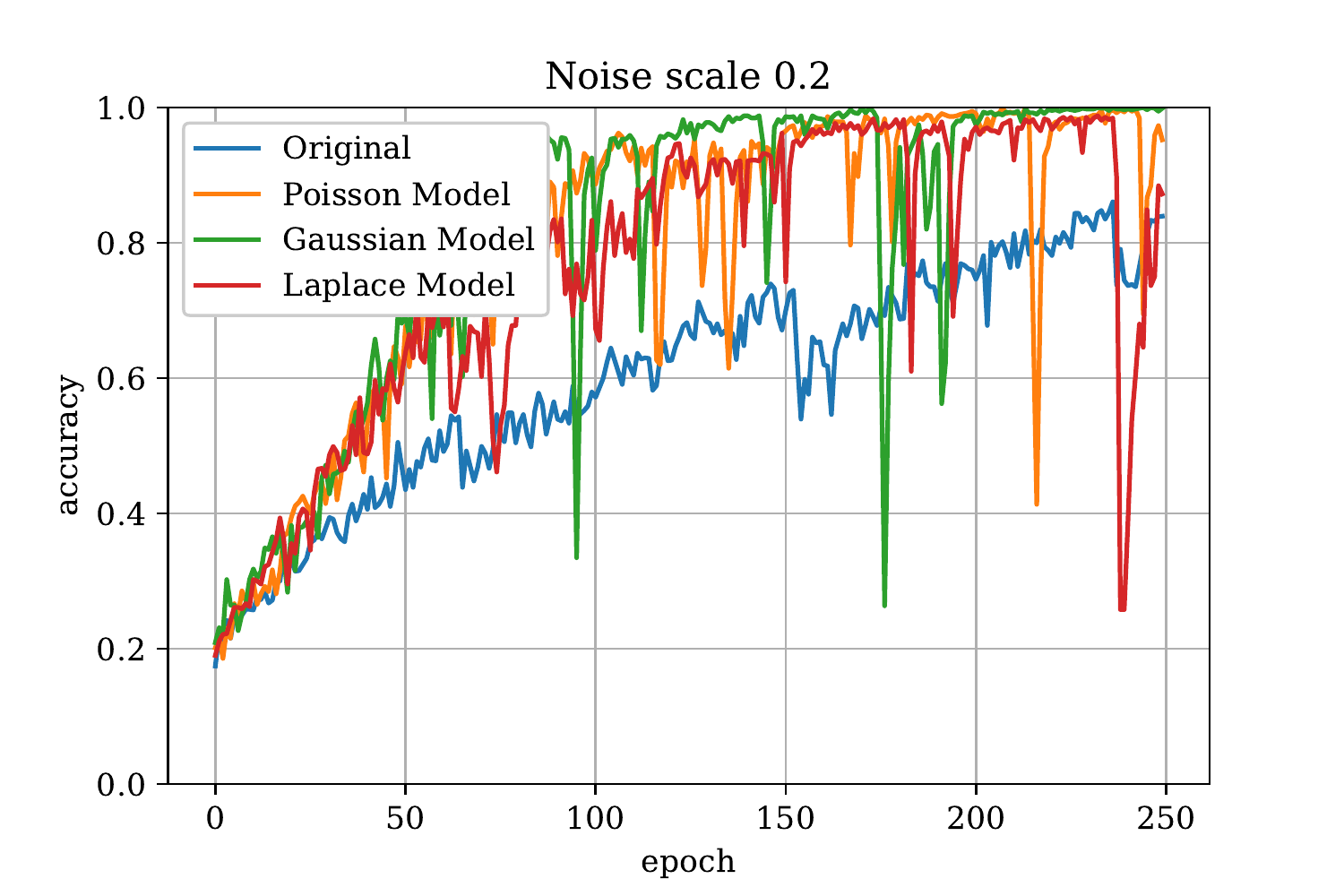}
		\label{fig:acc-plt-0.2}
	\end{subfigure}\hfill
	\begin{subfigure}{0.49\linewidth}
		\includegraphics[width=\linewidth]{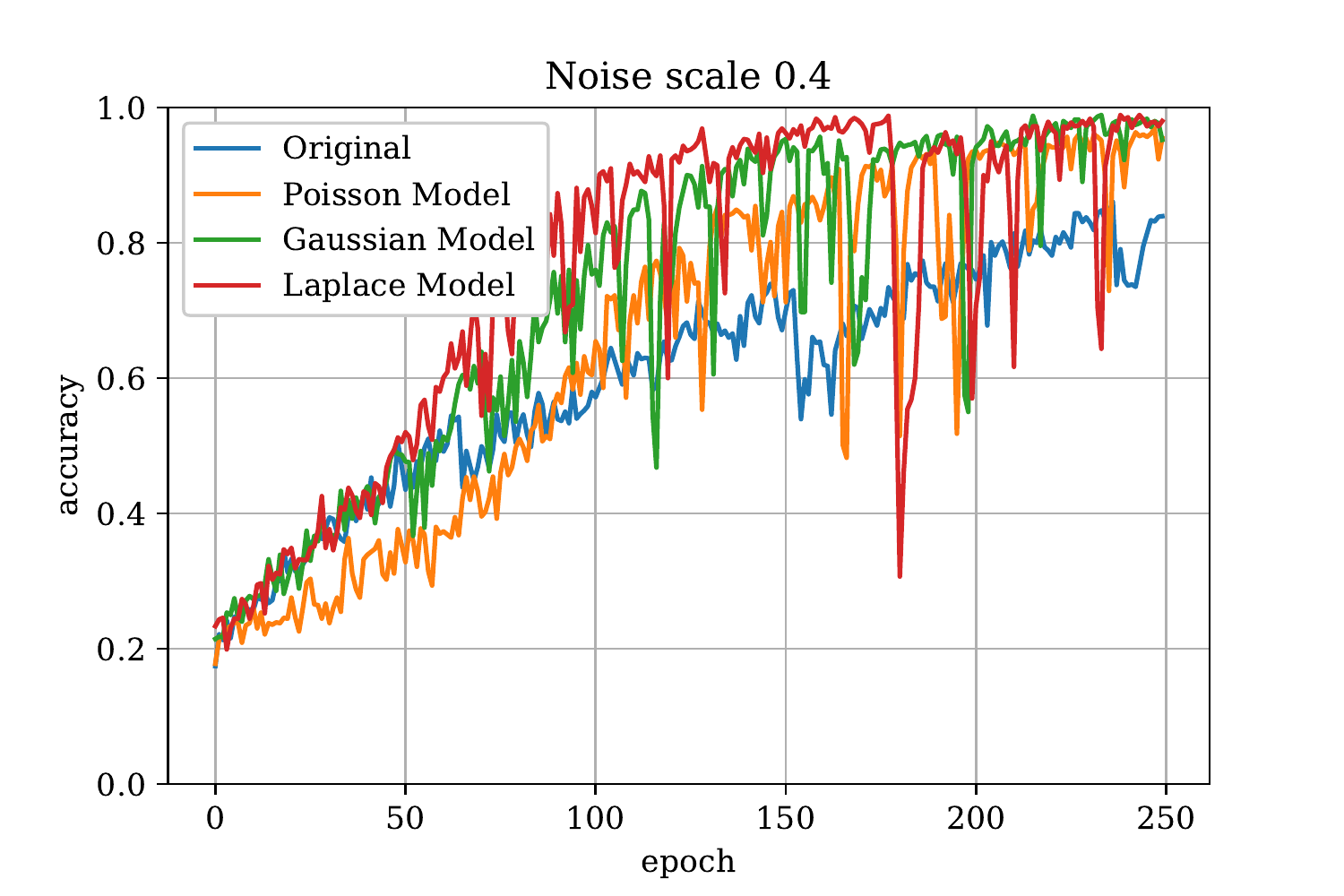}
		\label{fig:acc-plt-0.4}
	\end{subfigure}\hfill
	\begin{subfigure}{0.49\linewidth}
		\includegraphics[width=\linewidth]{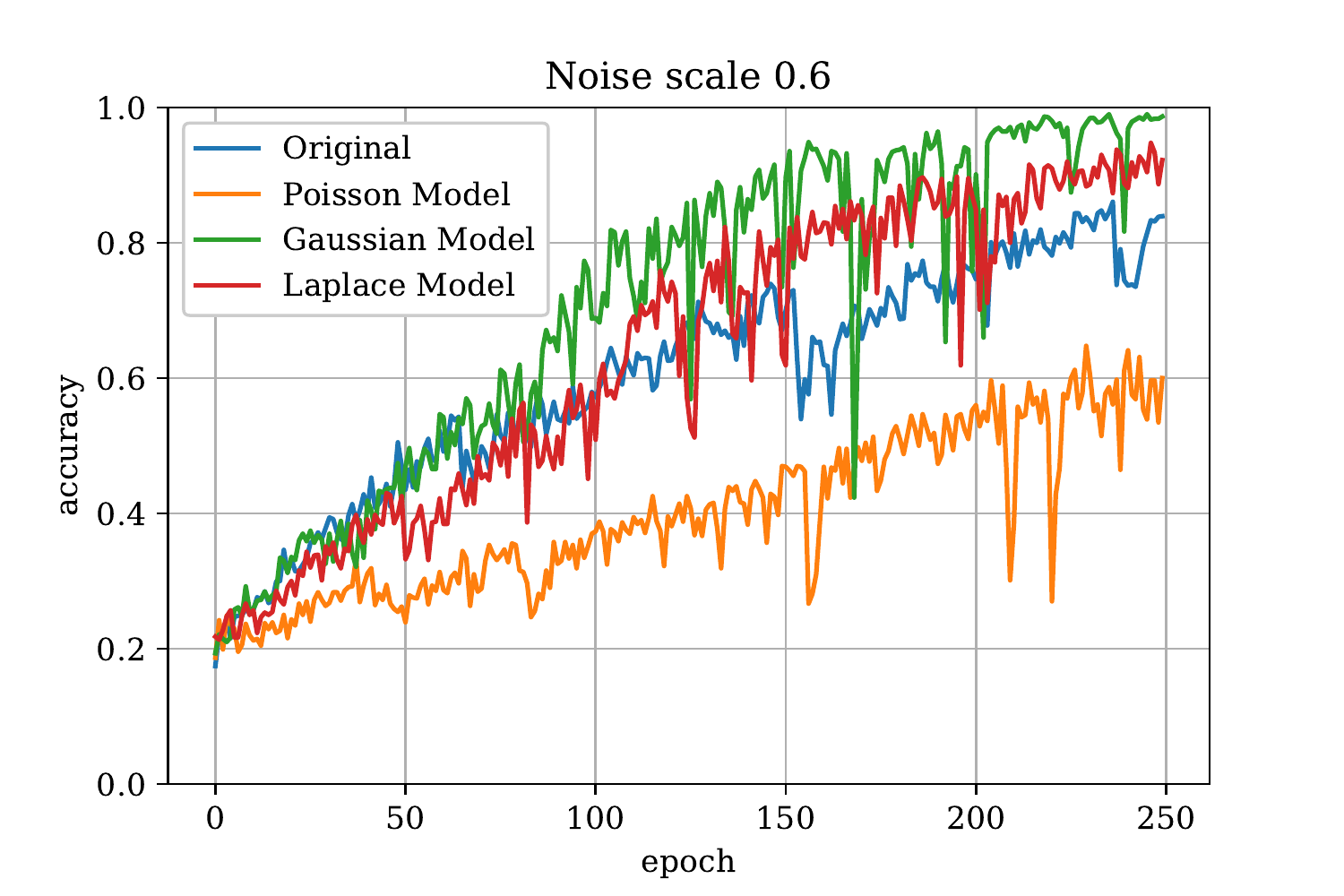}
		\label{fig:acc-plt-0.6}
	\end{subfigure}
	\begin{subfigure}{0.49\linewidth}
		\includegraphics[width=\linewidth]{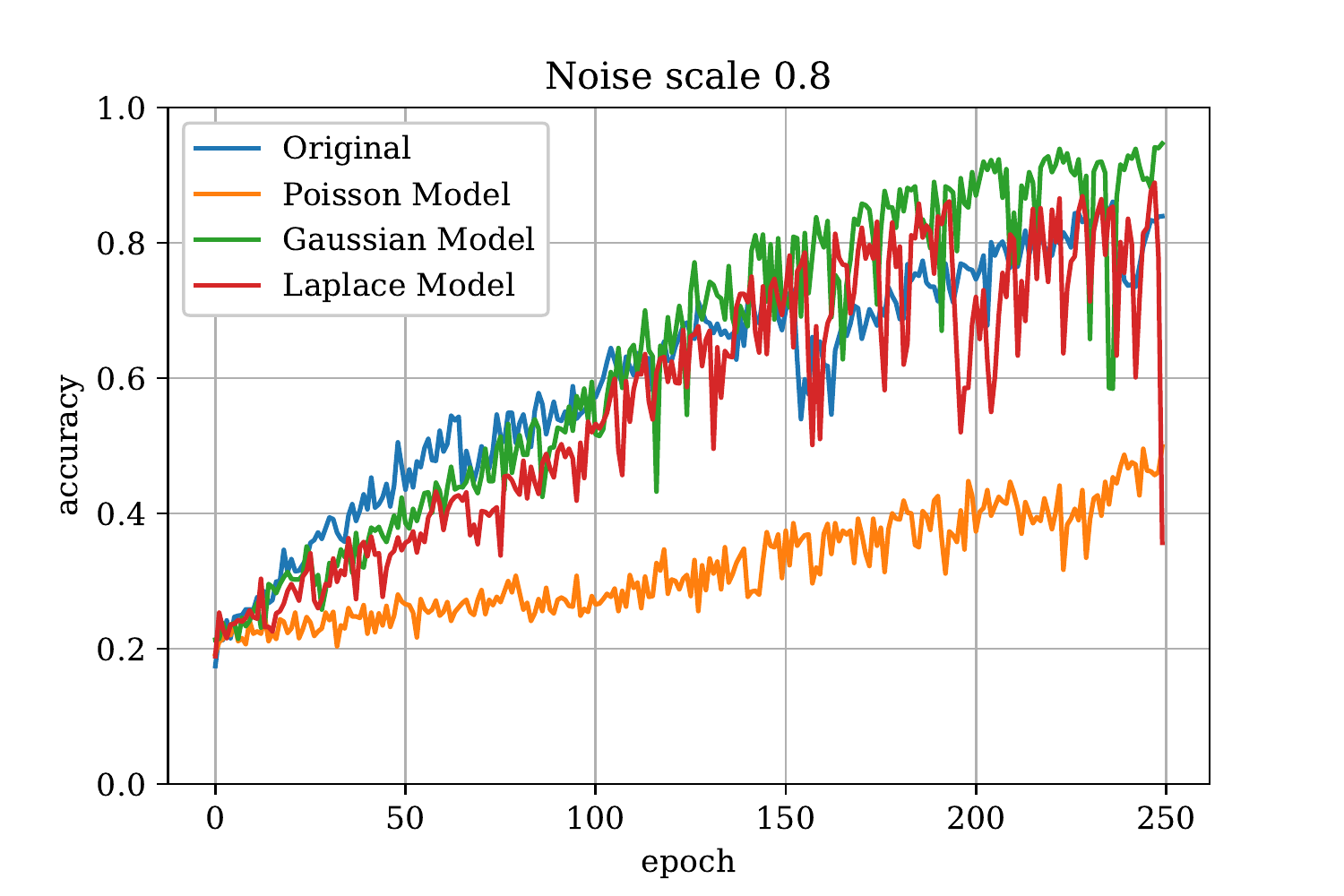}
		\label{fig:acc-plt-0.8}
	\end{subfigure}
	\begin{subfigure}{0.49\linewidth}
		\includegraphics[width=\linewidth]{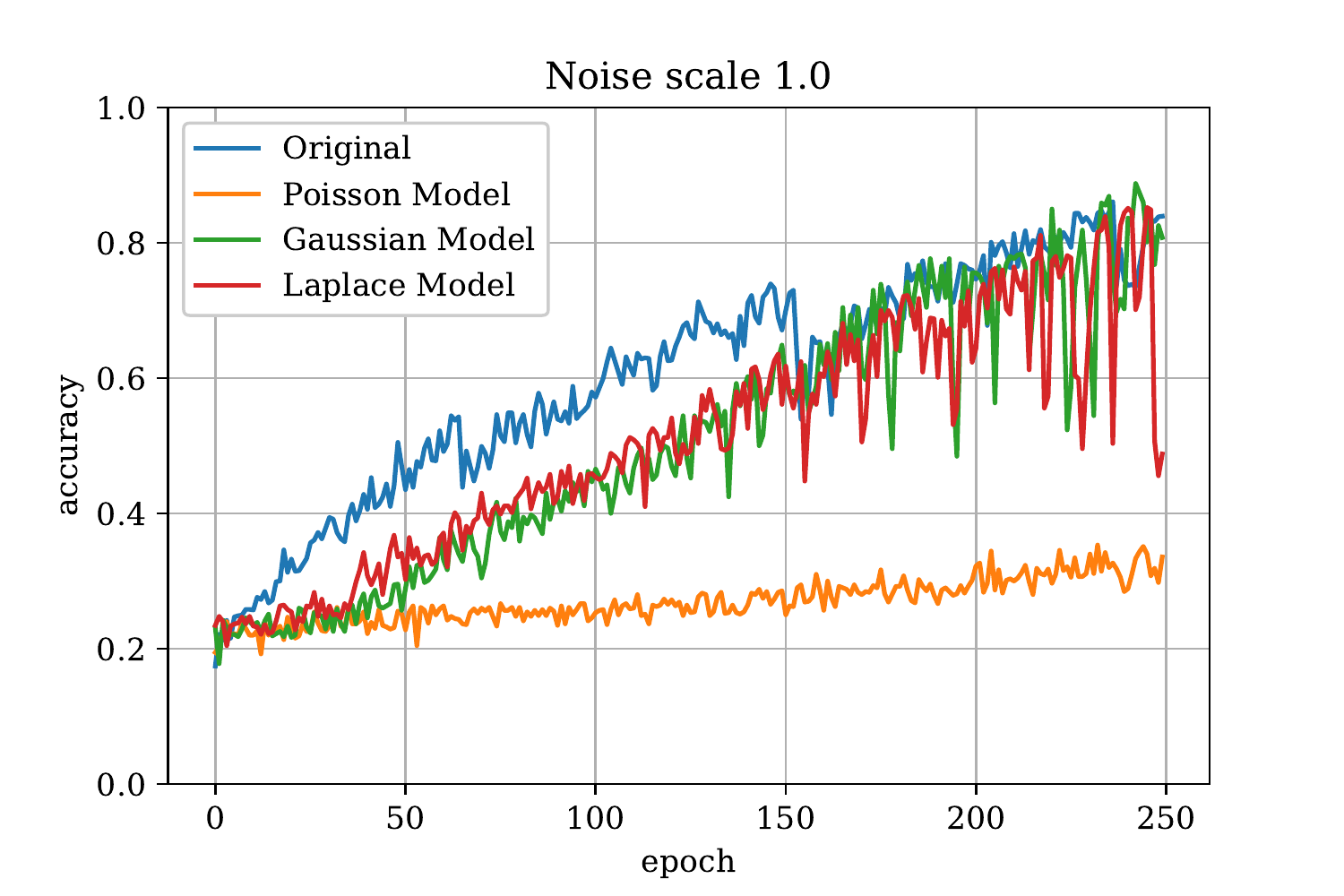}
		\label{fig:acc-plt-1.0}
	\end{subfigure}
	\caption{The different noise ratio accuracy results for additive Laplace/Gaussian/Poisson and original CNN model's accuracy results.}
	\label{fig:acc-plt-noise}
\end{figure}

Figure \ref{fig:acc-plt-noise} shows the accuracy change in each iteration of the CNN model, which is trained with the malware dataset containing a different amount of noise. The performance results of four CNN models, whose dataset is enriched by using both Additive Laplace, Additive Gaussian, and Additive Poisson methods, are better than the CNN model's classification performance that is trained only with the original training data set. When the noise ratio is 0.5, the original CNN model's classification result is better than the CNN model with the Additive Poisson method. When the noise ratio is increased to 0.8, the classification results of CNN models with Additive Gaussian, Additive Laplace, and Additive Poisson begin to decrease.

\begin{figure}[htbp!]
	\centering
	\begin{subfigure}{0.49\linewidth}
		\includegraphics[width=\linewidth]{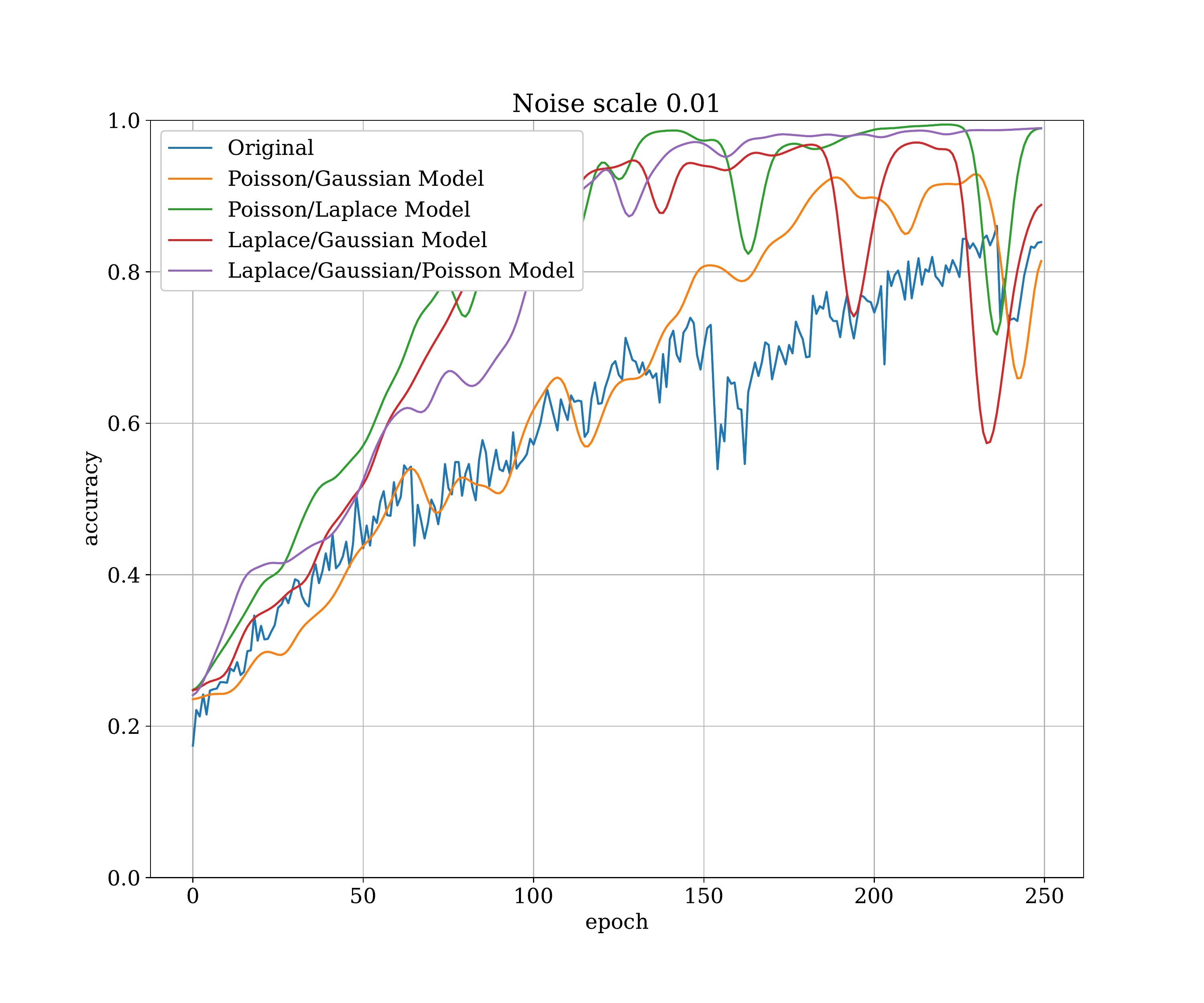} 
		\label{fig:acc-plt-comb-0.01}
	\end{subfigure}\hfill
	\begin{subfigure}{0.49\linewidth}
		\includegraphics[width=\linewidth]{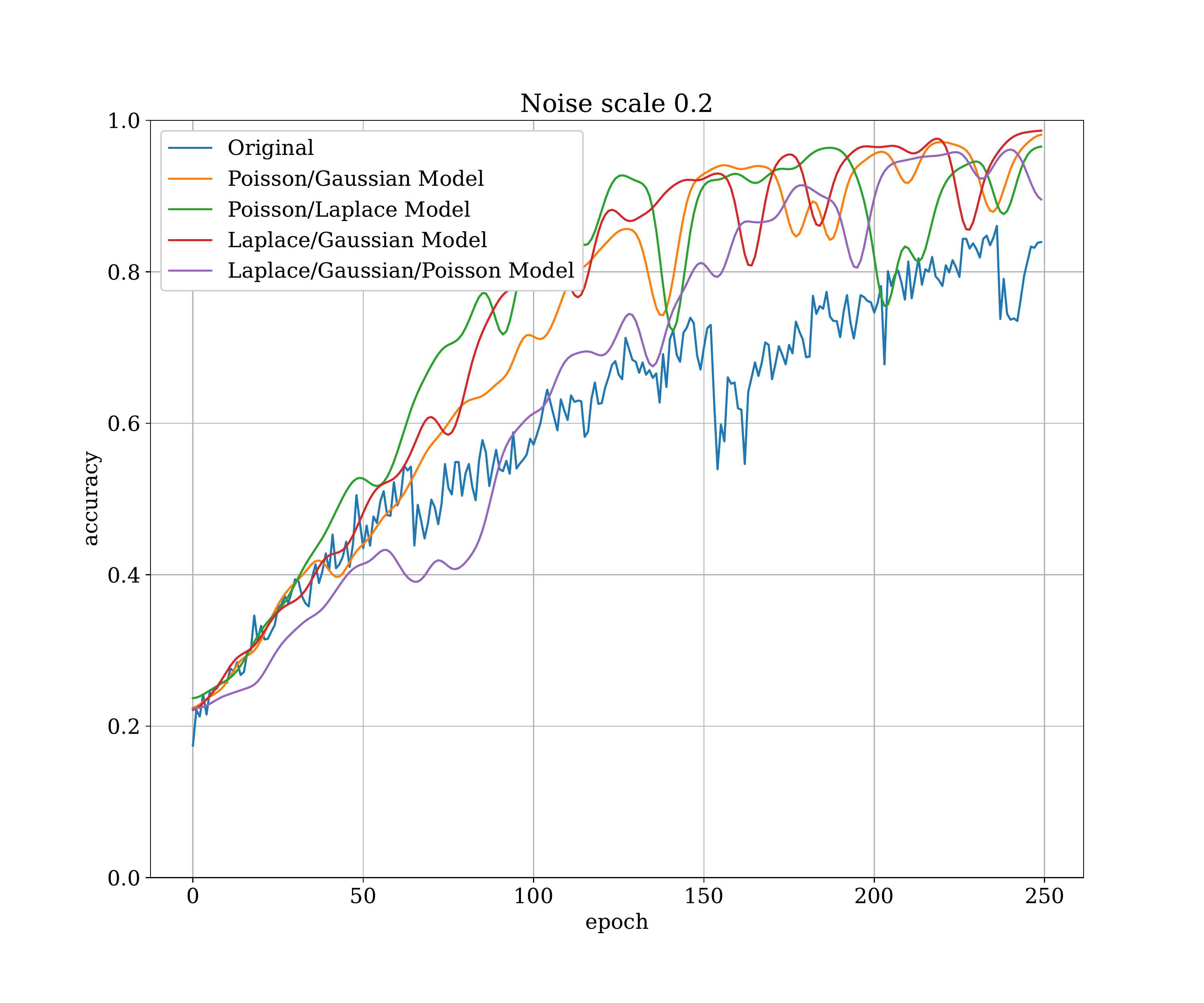}
		\label{fig:acc-plt-comb-0.2}
	\end{subfigure}\hfill
	\begin{subfigure}{0.49\linewidth}
		\includegraphics[width=\linewidth]{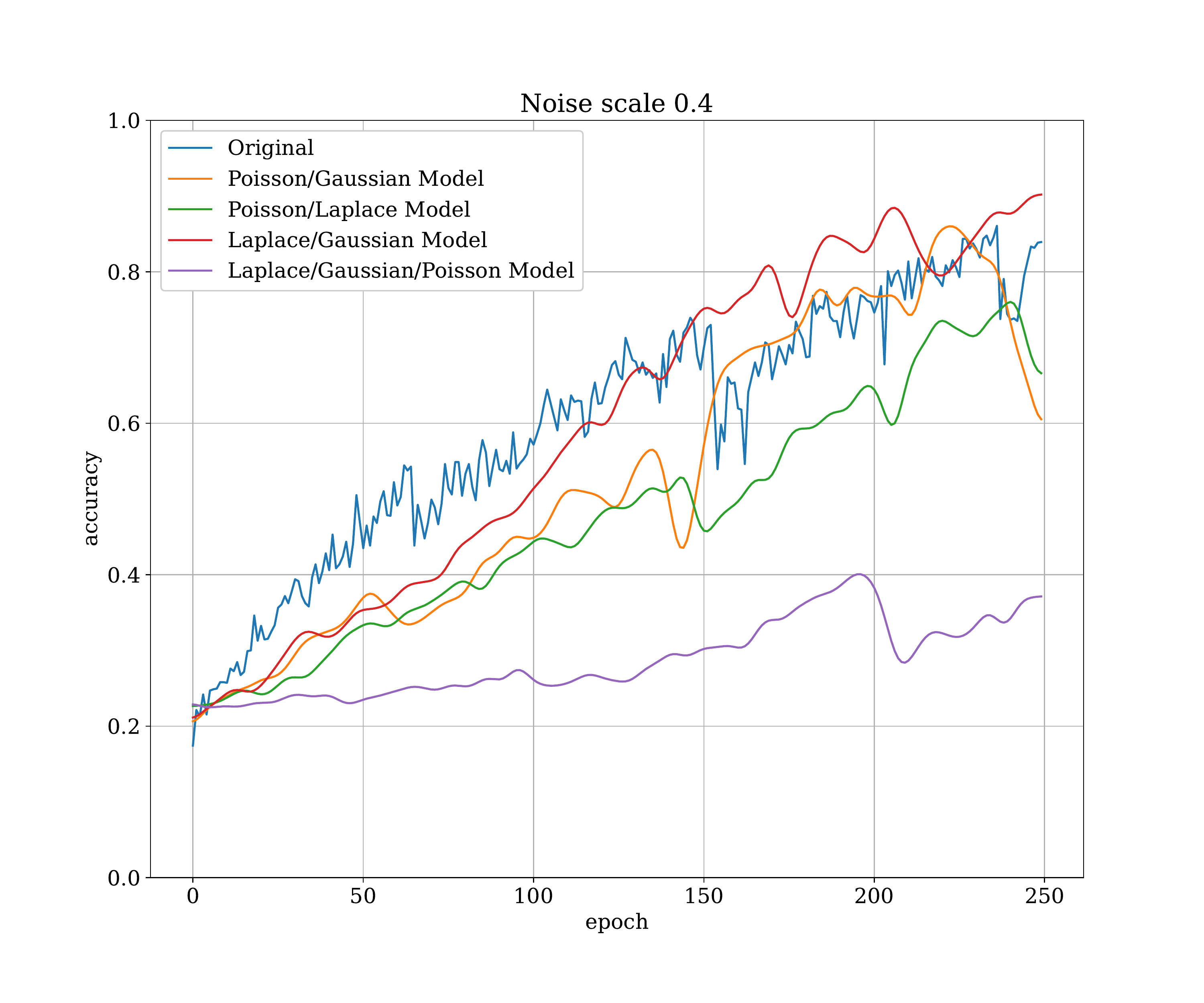}
		\label{fig:acc-plt-comb-0.4}
	\end{subfigure}\hfill
	\begin{subfigure}{0.49\linewidth}
		\includegraphics[width=\linewidth]{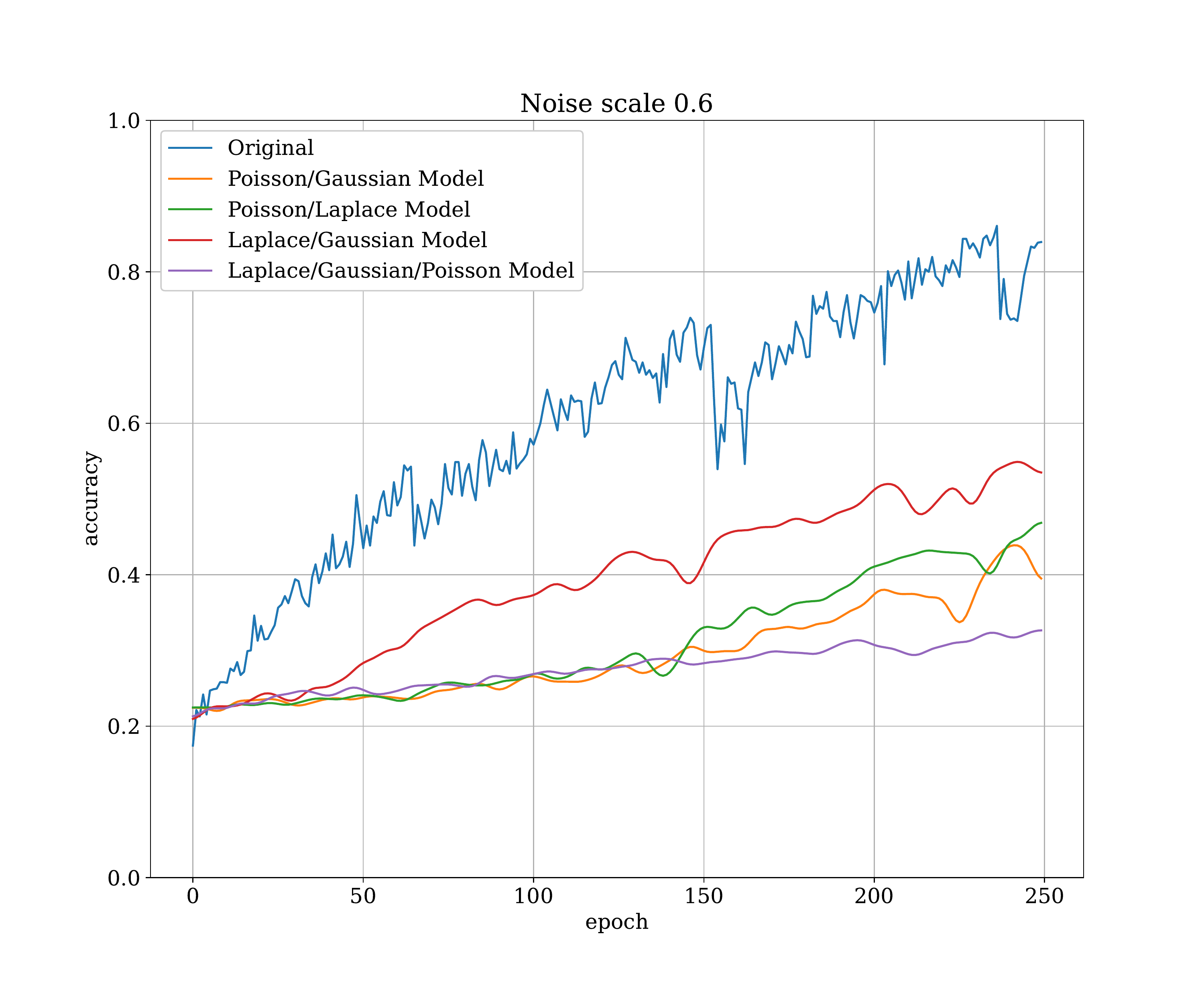}
		\label{fig:acc-plt-comb-0.6}
	\end{subfigure}
	\begin{subfigure}{0.49\linewidth}
		\includegraphics[width=\linewidth]{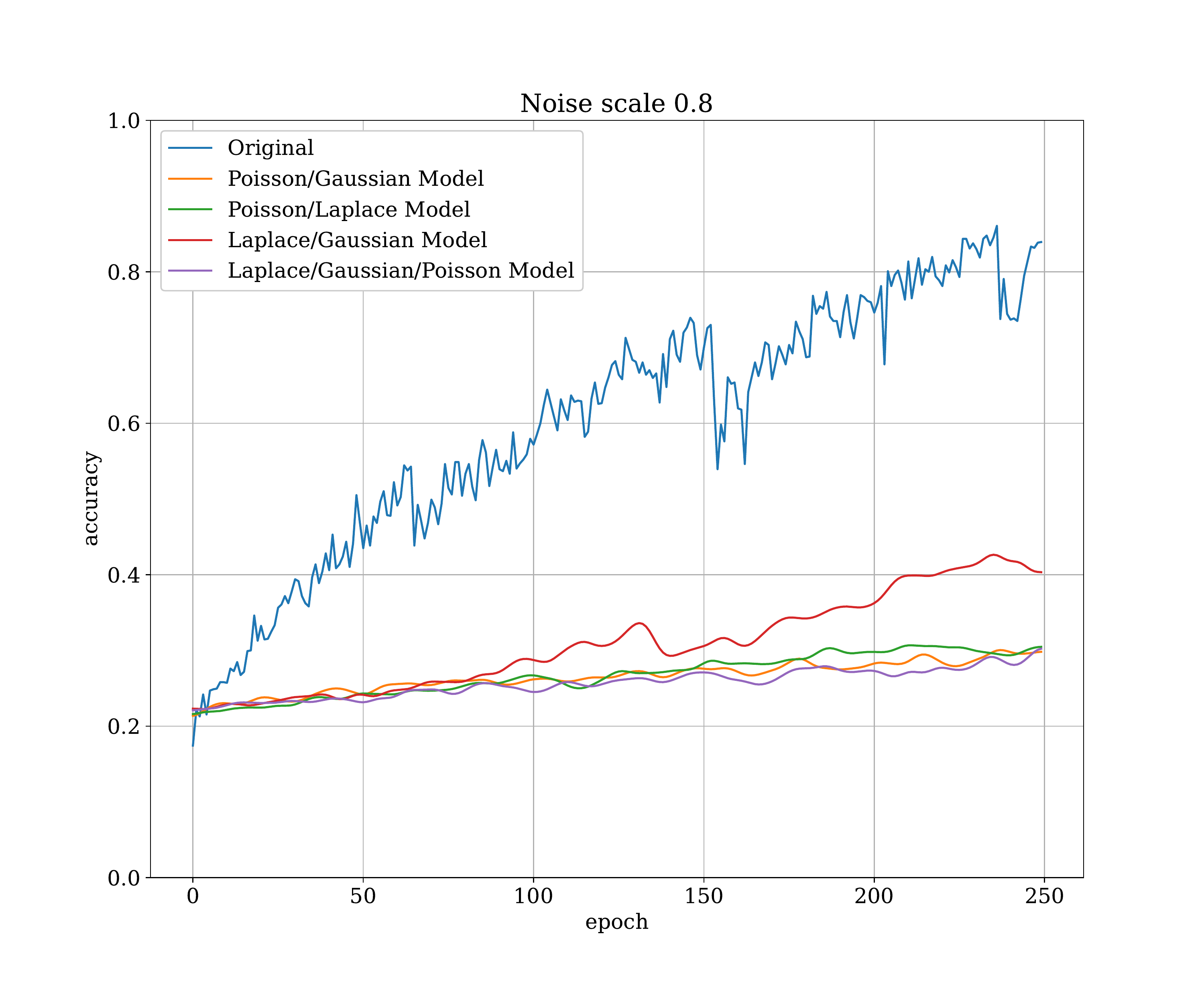}
		\label{fig:acc-plt-comb-0.8}
	\end{subfigure}
	\begin{subfigure}{0.49\linewidth}
		\includegraphics[width=\linewidth]{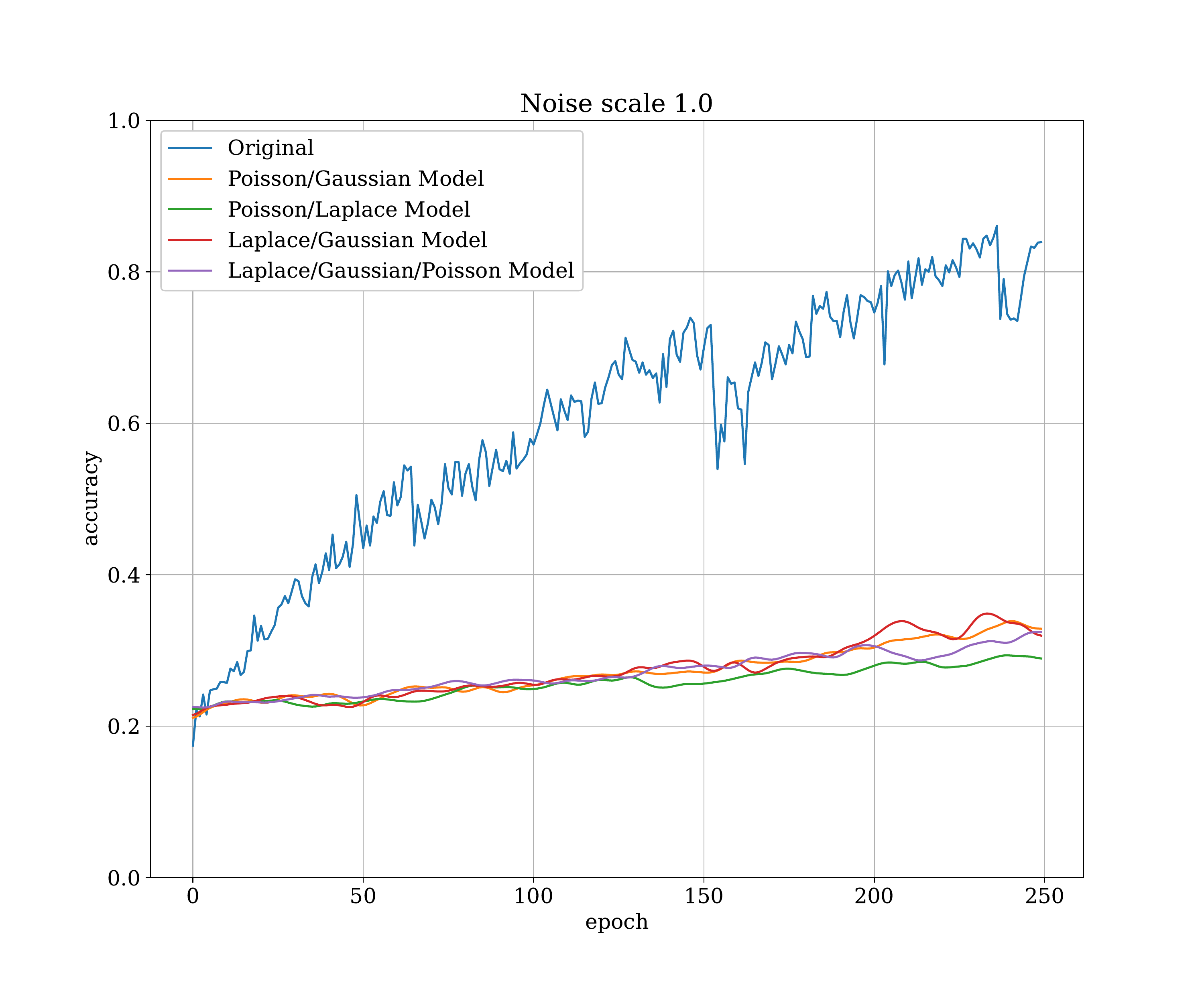}
		\label{fig:acc-plt-comb-1.0}
	\end{subfigure}
	\caption{The different noise ratio accuracy results for the combination of additive Laplace/Gaussian/Poisson and original CNN model's accuracy results.}
	\label{fig:acc-plt-noise-comb}
\end{figure}

Figure \ref{fig:acc-plt-noise-comb} shows the accuracy change in each iteration of the CNN model, which is trained with the malware dataset containing a different amount of noise with different combination of noise models. The performance results of five CNN models, whose dataset is enriched by using combination of Additive Laplace, Additive Gaussian, and Additive Poisson methods, are better than the CNN model's classification performance that is trained only with the original training data set. When the noise ratio is 0.4, the original CNN model's classification result is better than the CNN model with the several combination of noise injection methods.

Table \ref{tbl:noise-acc} shows the accuracy changes with different noise methods and different noise ratio. The fields shown as bold on the table show the best accuracy value of the column. The best accuracy value for Poisson noise is obtained with 0.902 and 0.3 noise ratio, the best accuracy value for Gaussian noise is obtained with 0.922 and 0.4 noise ratio, and the best accuracy value for Laplace noise is obtained with 0.819 and 0.2 noise ratio. According to the table, we obtain the best classification performance with the Gaussian noise's 0.4 noise ratio.

\begin{table}[h]
	\caption{Noise injection accuracy results}
	\label{tbl:noise-acc}
	\setlength{\tabcolsep}{3pt}
	\centering
	\begin{tabular}{|c|c|c|c|c|}
		\hline
		\textbf{Noise ratio} & \textbf{Orginal Model} & \textbf{Poission} & \textbf{Gaussian} & \textbf{Laplace} \\
		\hline
		0.01 & 0.83 & \textbf{1.00} & 0.96 & \textbf{0.99} \\ 
		0.2 & 0.83 & 0.95 & \textbf{0.99} & 0.87 \\ 
		0.4 & 0.83 & 0.95 & 0.95 & 0.98 \\ 
		0.6 & 0.83 & 0.60 & 0.98 & 0.92 \\ 
		0.8 & 0.83 & 0.49 & 0.94 & 0.35 \\ 
		1.0 & 0.83 & 0.33 & 0.80 & 0.48 \\\hline
	\end{tabular}
\end{table}

Table \ref{tbl:noise-acc-comb} shows the accuracy changes with the different combination of noise methods and different noise ratio. The fields shown as bold on the table show the best accuracy value of the column. The best accuracy value for Poisson/Gaussian noise is obtained with 0.93 and 0.2 noise ratio, the best accuracy value for Poisson/Laplace noise is obtained with 0.95 and 0.01 noise ratio, the best accuracy value for Laplace/Gaussian noise is obtained with 1.00 and 0.01 noise ratio. 

\begin{table}[h]
	\centering
	\caption{The best accuracy rates for the combination of each noise type}
	\label{tbl:noise-acc-comb}
	\setlength{\tabcolsep}{3pt}
	\begin{tabular}{|c|c|p{1.2cm}|p{1.2cm}|p{1.2cm}|c|}
		\hline
		\textbf{Noise} & \textbf{Org} & \textbf{Poisson / Gaussian} & \textbf{Poisson / Laplace} & \textbf{Laplace / Gaussian} & \textbf{All} \\
		\hline
		0.01 & 0.83 & 0.90 & \textbf{0.95} & \textbf{0.98} & \textbf{0.96}\\ 
		0.2  & 0.83 & \textbf{0.93} & 0.90 & 0.95 & 0.95\\ 
		0.4  & 0.83 & 0.90 & 0.71 & 0.90 & 0.42\\ 
		0.6  & 0.83 & 0.47 & 0.52 & 0.38 & 0.76\\ 
		0.8  & 0.83 & 0.52 & 0.47 & 0.76 & 0.66\\ 
		1.0  & 0.83 & 0.76 & 0.52 & 0.47 & 0.76\\\hline
	\end{tabular}
\end{table}

\begin{figure}[h]
	\centering
	\includegraphics[width=0.5\linewidth]{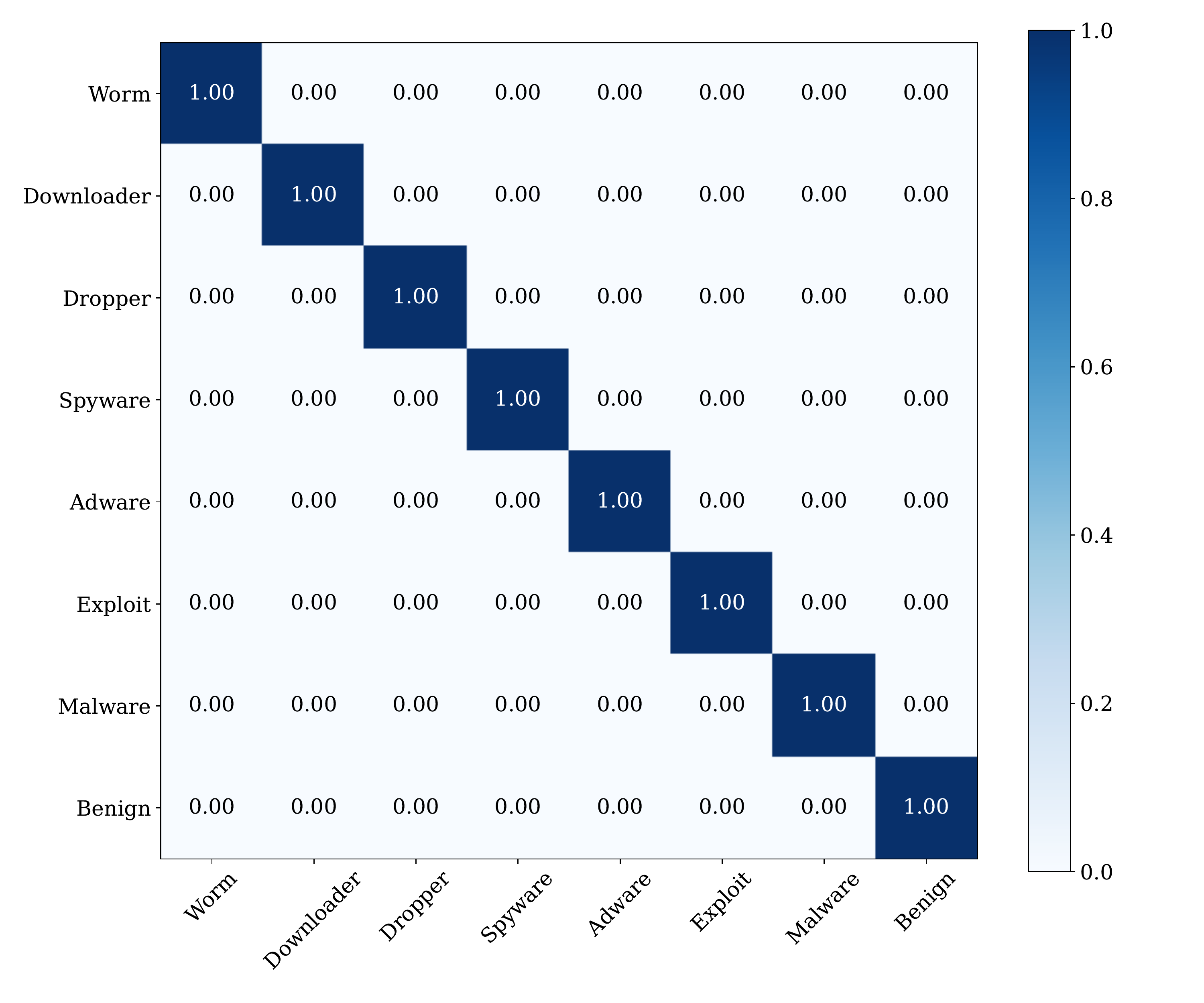}
	\caption{The confusion matrix of the CNN model with best data noise injection ratio.}
	\label{fig:aug-conf-mat-best}
\end{figure}

The best classification performance is performed by using the Poisson noise with 0.01 value has a 100\% classification performance. Figure \ref{fig:aug-conf-mat-best} shows the confusion matrix of the malware detection model with the best classification performance. 

\section{Conclusion}\label{sec:conclusion}
The research study's primary purpose is to provide an image augmentation enhanced deep convolutional neural network (CNN) model for the detection of malware families in a metamorphic malware environment. The model architecture consists of three main components: image generation from malware samples, image augmentation, and classifying the malware families by using convolutional neural network models. In the first component, the collected malware samples are converted into binary representation using the windowing technique. We apply augmentation techniques in the second component using the imgaug Python library. We enhanced our dataset using additive noise techniques such as Gaussian, Laplacian, and Poisson. To demonstrate the proposed model's effectiveness and classification performance, we apply it to our dataset with different hyper-parameters. Finally, we train our classifier on our public dataset with seven different classes, including Worm, Downloader, Spyware, Adware, Exploit, Malware, and 346 Benign. The model becomes its steady-state after the 200th epoch.

We observe that the training dataset shows more stable progress, the test dataset is less stable, although it progresses together. We use four different metrics, the overall prediction accuracy, average recall, average precision, and F1-score, to evaluate the classification accuracy. The confusion matrix results indicate that the classification model performance is not good enough for malware detection. The classification performance of the model obtained with conventional CNN is relatively low. According to these results, a standard CNN model with an RGB type 3-channel image training dataset is not suitable for malware detection and classification. The accuracy results of train and test datasets over the proposed method on the augmented dataset. The augmentation is measured with varying noise ratio to assess the effectiveness of the learning algorithm. This paper's main contribution is to propose a data augmentation enhanced malware family classification model that exploits augmentation for variants of malware clones and takes advantage of CNN to improve image classification. It is evident from the results of this research that the data augmentation based on 3-channel image classification can significantly influence the performance of malware family classification.

 \bibliographystyle{acm}
\bibliography{mybibfile}

\end{document}